\documentclass[a4paper,fleqn,usenatbib]{mnras}

\usepackage{natbib}
\usepackage{multicol}

\usepackage{graphicx}	
\usepackage{amsmath}	
\usepackage{amssymb}	
\usepackage{subcaption}
\usepackage{float}
\usepackage{multirow}
\usepackage{bm}
\usepackage{mathabx}

\usepackage{algorithm,algcompatible,amsmath}

\DeclareGraphicsExtensions{.pdf,.png,.jpg,.eps}
\graphicspath{{figures/}}

\newcommand{\mean}[1]{\ensuremath{\left\langle #1 \right\rangle}}

\def\reff@jnl#1{{\rm#1\/}}

\def\aj{\reff@jnl{AJ}}                  
\def\araa{\reff@jnl{ARA\&A}}            
\def\apj{\reff@jnl{ApJ}}                        
\def\apjl{\reff@jnl{ApJ}}               
\def\apjs{\reff@jnl{ApJS}}              
\def\apss{\reff@jnl{Ap\&SS}}            
\def\aap{\reff@jnl{A\&A}}               
\def\aapr{\reff@jnl{A\&A~Rev.}}         
\def\aaps{\reff@jnl{A\&AS}}             
\def\baas{\reff@jnl{BAAS}}              
\def\jrasc{\reff@jnl{JRASC}}            
\def\memras{\reff@jnl{MmRAS}}           
\def\mnras{\reff@jnl{MNRAS}}            
\def\physrep{\reff@jnl{Phys.Rep.}}
\def\pra{\reff@jnl{Phys.Rev.A}}         
\def\prb{\reff@jnl{Phys.Rev.B}}         
\def\prc{\reff@jnl{Phys.Rev.C}}         
\def\prd{\reff@jnl{Phys.Rev.D}}         
\def\prl{\reff@jnl{Phys.Rev.Lett}}      
\def\pasp{\reff@jnl{PASP}}              
\def\pasj{\reff@jnl{PASJ}}              
\def\skytel{\reff@jnl{S\&T}}            
\def\solphys{\reff@jnl{Solar~Phys.}}    
\def\sovast{\reff@jnl{Soviet~Ast.}}     
\def\ssr{\reff@jnl{Space~Sci.Rev.}}     
\def\nat{\reff@jnl{Nature}}             

\newcommand{\beq}{\begin{equation}}
\newcommand{\eeq}{\end{equation}}
\newcommand{\beqa}{\begin{eqnarray}}
\newcommand{\eeqa}{\end{eqnarray}}

\usepackage{color}
\definecolor{sdeep}{rgb}{0., 0.6, 0.}
\newcommand{\referee}[1]{{{ #1}}}

\definecolor{english}{rgb}{0.0, 0.5, 0.0}
\newcommand{\refereetwo}[1]{{{ #1}}}

\newcommand{\vx}{\ensuremath{\bm {x}}}
\newcommand{\wj}[6]{ \begin{pmatrix} 
   #1 & #2 & #3 \\
   #4 & #5 & #6 
  \end{pmatrix}}
  
\newcommand{\vtheta}{{\bm \theta}}

\title[\sc $i$Master]{$i$mproved {\sc Master} for the LSS: Fast and accurate analysis of the two point power spectra and correlation functions
}

\author[S.~Singh ]{
   Sukhdeep Singh$^{1,2}$\thanks{E-mail: sukhdeep@cmu.edu}
\\
$^1$McWilliams Center for Cosmology, Department of Physics, Carnegie Mellon University, Pittsburgh, PA 15213, USA\\
$^2$Berkeley Center for Cosmological Physics, University of California, Berkeley, CA 94720, USA\\
}

\date{Accepted XXX. Received YYY; in original form ZZZ}

\pubyear{2021}

\begin{document}
\label{firstpage}
\pagerange{\pageref{firstpage}--\pageref{lastpage}}
\maketitle

\begin{abstract}
We review the methodology for measurements of two point functions of the cosmological observables, both power spectra and correlation functions.
For pseudo-$C_\ell$ estimators, we will argue that the  window weighted overdensity field can yield more optimal measurements as the window acts as an inverse 
noise weight, an effect that becomes more important for surveys with a variable selection function. 
We then discuss the impact of approximations made in the {\sc Master} algorithm and suggest improvements, the {\sc $i$Master} algorithm, that uses the theoretical model to give unbiased results 
for arbitrarily complex windows provided that the model satisfies weak accuracy conditions. 
The methodology of {\sc $i$Master} algorithm is also generalized to the correlation functions to reconstruct the binned power spectra, for E/B mode separation, or to properly \refereetwo{convolve} the correlation functions 
to account for the scale cuts in the Fourier space model.
We also show that the errors in the window estimation lead to both additive and multiplicative effects on the over density field.  
Accurate estimation of window power can be required up to scales of $\sim 2\ell_\text{max}$ or larger. Misestimation of the window power leads to biases in the measured power spectra which scale 
as ${\delta C_\ell}\sim M^W_{\ell\ell'}\delta W_{\ell'}$, where the $M^W_{\ell\ell'}$ scales as $\sim(2\ell+1)C_\ell$ leading to effects that can be important at high $\ell$.
While the notation in this paper is geared towards photometric galaxy surveys, the discussion is equally applicable to spectroscopic galaxy, intensity mapping and CMB surveys.
\end{abstract}




\section{Introduction}

The measurements of the large scale structure (LSS) in the universe provide important cosmological information about the evolution of the universe over time and also allow us to study the 
physical properties of its constituents, namely Dark matter, Dark energy, neutrinos and baryons \citep[see ][ for a review]{Weinberg2013}. Over the past two decades we have successfully measured the LSS using a number of probes, e.g. baryon acoustic oscillations and galaxy velocities \citep[e.g.][]{Boss2016combined,Neveux2020}, weak gravitational lensing\citep[e.g.][]{Planck2018lensing,DES2017comb,Singh2020cosmo,Heymans2021}, galaxy clusters \citep[e.g.][]{DESY1_cluster}. 
These measurements have yielded strong constraints on the cosmological models with precision of order 5-10\%.
There are also some tensions among the probes at $2-3\sigma$ significance level \citep{DES2017comb,Singh2020cosmo,Heymans2021,Valentino2020,Lange2021} that have been a source of general intrigue and 
excitement in the community and also require us to revisit 
many of the assumptions made in the analysis. With upcoming percent level measurements from the planned surveys, e.g. Rubin Observatory LSST-DESC \citep{LSST_DESC}, DESI \citep{DESI2019}, Roman 
space telescope \citep{Wfirst2019}, Simons Observatory \citep{SO2019}, CMB-S4 \citep{CMB_S4}, Spherex \citep{Spherex2014}, it is going to become even more important for us to thoroughly understand the 
analysis in order to derive accurate inferences on cosmological models. In this paper we will review the methodology for the measurements of the two point statistics of the over density field from the LSS surveys.

The large scale structure measurements from the cosmological surveys are made by turning the data from the survey into the maps of over density field, which is the quantity of interest in measuring the 
fluctuations in the matter density. Correlations of mean zero overdensity field have also been shown to be more optimal than the density field with non-zero mean \citep{Landy1993,Singh2017cov}. \referee{In this paper, by optimal we will usually mean minimizing some combination of bias and variance in 
the inferences.}
The maps of over density are then compressed using the summary statistics, especially the two point correlation functions or power spectra, which are a measure of the variance in the 
field. During the process of these measurements there are several questions that one needs to address to ensure the optimality of the measurement as well as the accuracy of the inferences derived from these measurements.

After the data is acquired and cleaned, the analysis begins with the process of map making (or making catalogs), 
whereby we bin the count of galaxies (or photons in case of intensity mapping) into pixels of a map. One of the 
questions that we need to address at this stage is the weighting to be applied to the pixels to ensure optimal analysis. The question of optimal weights for the case of gaussian random fields has been addressed by \cite{FKP} giving us the famous FKP weights, which was later generalized by \cite{Hamilton1997}. While 
FKP weights (or quadratic estimator in general) has been shown to be optimal for the gaussian field, 
it is computationally expensive to use such weights in practice and instead the pseudo-$C_\ell$ like estimators 
\citep{Wandelt2001} are preferred. In such estimators, it is fairly common to adopt uniform weighting, which is usually not optimal. 

The next step in the analysis is to measure the power spectra or the correlation functions of the over density field. In this step, one of the major challenges is to account for the effects of the survey 
window. The survey window or the selection function depends on the survey geometry (mask) as well as the any observational selection effects and the weights applied on the maps
as discussed earlier. The modeling of window effects on the pseudo-$C_\ell$ power spectra has been addressed in detail by \cite{Hivon2002} who introduced the now standard 
{\sc Master} algorithm to deconvolve 
the effects of the window to reconstruct the power spectra of the underlying fields. Similarly, formalism for the measurement of the correlation functions was developed by 
\cite{Ng1999}. The window also affects the correlation functions in a very similar manner by acting as a weighting function and changing the effective scale of measurement
 \citep[see e.g. appendix D of][]{Singh2020cosmo}. 

Since proper modeling of the two point functions requires careful modeling of the window effects, one of the major challenges in the LSS measurements is to properly estimate the window. Biases in the 
window estimation directly propagate into both the power spectra and the correlation functions. The strategies to model window biases include the mode deprojection \citep{Slosar2004,Leistedt2013,Elsner2017}
where one can subtract out systematics or down weight the affected modes by including additional terms in the covariance. Similar strategies have also been applied for the case of the 
correlation functions, e.g. \cite{Ross2012}.

The final step before running the inference is to implement the `scale cuts' on the measurements to match the scales of the model. 
Typically the models we use to analyze the data are validated 
under certain assumptions and can only model a limited range of scales which is smaller than the range of scales probed by the current data. 
Applying these scale cuts carefully is necessary for the optimal 
analysis, i.e. to extract as much information as possible while avoiding the biases from the model outside its range of validity. Recently this issue of scale cuts has shown up in 
some apparent discrepancies
between the analysis in the configuration space (correlation functions) and the Fourier space (power spectra) \citep[see ][]{Hamana2020,Doux2021}. Such issues are concerning
as they complicate the interpretation of the inferences drawn from the analysis and need to be carefully addressed.

In this paper we will review the methodology described above, study the optimality of the weights applied on the over density field during the pseudo-$C_\ell$ like analysis and also
test the assumptions made in the methods to model the window functions. Since modeling the window function is important and also very challenging, we will study the impact of the window mis-estimation
on the two point functions (\refereetwo{we will not discuss the methods to estimate the window, only the methods to account for window mis-estimation}). Finally, we will generalize the {\sc Master} algorithm to the correlation functions to study the impact of the scale cuts and devise new methods for more optimal scale cuts.
We will use the notation commonly used in the analysis of the angular statistics. However, most of our discussion will be equally applicable to the analysis of the spectroscopic surveys, both 
galaxies and intensity mapping ones.

The simulations and power spectra calculations for window modeling are performed using the {\sc Healpy}\citep{Gorski2005} package, the matter power spectra is obtained using {\sc Camb}\footnote{\url{http://
camb.info}} package and the 
computations of angular power spectra, coupling matrices, correlation functions and any other relevant calculations are performed using the {\sc Skylens} package (Singh et al. in prep).

	\section{Map making}
	\label{sec:maps}
		We begin with a brief the discussion of turning the catalogs of observables from the telescopes to the maps of the over density fields.
		For the case of galaxy counts ( or intensity mapping), the observed number of galaxies at a position $\vx$ can be described as
		\begin{align}
			{n}_{g}(\vx)=\mean{{n}_g(\vx)}(1+\delta_g(\vx)),
			\label{eq:n_gx}
		\end{align}
		where $\mean{n_g(\vx)}$ is the ensemble average of $n_g(\vx)$ and can be thought of as the expected number of galaxies to be observed at position $\vx$ ignoring the effects of noise and 
		the over density field. 
		$\delta_g(\vx)$ is the 
		overdensity of galaxies at $\vx$. 
		We will also use the mean of $\widebar{n}_g(\vx)$ over the whole survey,
		\begin{align}
			\widebar{n}_g=\widebar{n}_g(\vx).
			\label{eq:n_g}
		\end{align}
		Throughout this paper, $\mean{}$ implies the ensemble average of the quantity inside the brackets, \referee{which implies an average over many realizations of data} and $\widebar X$ 
		represents the sample mean of quantity $X$, \referee{where by sample mean we will usually imply mean over the given realization of the survey. Quantities with tophat, e.g. $
		\widehat\delta$, represent the measurements from data.}
		
		From eq.~\eqref{eq:n_gx}, a simple way to define windowed over density field is
		\begin{align}
			\widehat{\delta}_{g,W}(\vx)=\frac{\mean{n_g(\vx)}}{\widebar{n}_g}(1+\delta_g(\vx))-\frac{\mean{n_g(\vx)}}{\widebar{n}_g}=W_g(\vx)\delta_g(\vx),
			\label{eq:Pdelta_g}
		\end{align}
		In the second equality, we defined the window function (or selection function) as 
		\begin{align}
			W_g(\vx)=\frac{\mean{n_g(\vx)}}{\widebar{n}_g}=1+ F(C_i(\vx)).
			\label{eq:W_gF}
		\end{align}
		$W_g(\vx)$ is the selection function which accounts for the observational effects in the survey which modulate the observed density of the galaxies.
		In the second part of the equation we wrote the window as function of underlying contaminants, $C_i$. In general $F$ can be a non-linear function and a detailed discussion of window 
		estimation is outside the 
		scope of this work 
		\referee{\citep[see ][ for some recent work]{Ross2020,Rezaie2020,Everett2020}}. We will only focus on the effects of the window on statistics of interest, namely the two point correlations of the fields.
		We will discuss the effects of window miss-estimation in section~\ref{ssec:map_err} and in later sections on how it propagates into the two point correlations of the fields.
		In this work it is also assumed that the window function and the underlying over-density field are uncorrelated. This is in general not true as the 
		window function is correlated with the underlying field due to observational effects such as blending and fiber collisions. Detailed discussion of such effects is left for the future work.

		A more popular choice for defining over density maps is to remove the effects of window by dividing with $\mean{n_g(\vx)}$ (\refereetwo{or $W_g(\vx)$ as discussed above}), in which case 
		we get
		\begin{align}
			\widehat{\delta}_g(\vx)=\frac{{n_g(\vx)}}{\mean{n_g(\vx)}}-1=\frac{{n_g(\vx)}}{W_g(\vx)\widebar{n}_g}-1=\delta_g(\vx),
			\label{eq:delta_g}
		\end{align}
		which is free from the effects of window. This is not strictly true as the mask effects are still present and typically mask also needs to be modified to remove the
		pixels where $W_g(\vx)$ is small.
		 
		Many studies in the literature work with the galaxy catalogs instead of maps (see also discussion in section~\ref{ssec:LS_estimator}). 
		In such a case, if we apply systematics weights on galaxies to correct for window effects and use uniform randoms, \referee{i.e. ${\mean{n_R(\vx)}}/{\widebar{n}_g}=1$, where $n_R$ is the 
		number of randoms}, 
		\refereetwo{
		we get the estimator in eq.~\eqref{eq:delta_g} \citep[e.g.][]{Ross2012,Boss2016combined,Elvin-Poole2018}, i.e., 
		\begin{equation}
				\widehat{\delta}_{g}(\vx)=\frac{n_g(\vx)/W_g(\vx)-n_R(\vx)}{n_R(\vx)}=\frac{{n_g(\vx)}}{W_g(\vx)\widebar{n}_g}-1.
				\label{eq:delta_g_R}
			\end{equation}
		On the other hand, if the systematics weights are applied on the 
		randoms, \referee{i.e. ${\mean{n_R(\vx)}}/{\widebar{n}_g}=W_{g}(\vx)$}, then we obtain the estimator from eq.~\eqref{eq:Pdelta_g}, i.e., 
		\begin{equation}
			\widehat{\delta}_{g,W}(\vx)=\frac{n_g(\vx)-n_R(\vx)}{\widebar{n}_g}
			\label{eq:Pdelta_g_R}
		\end{equation}			
			}
		
		In eq.~\eqref{eq:Pdelta_g}, Window ($W_g$)  acts as weight on the pixels in the map and turns out to be a nearly optimal way to apply the weights. 
		In general, optimal weights for the power spectra estimation of a gaussian random field are \citep{FKP,Hamilton1997}
		\begin{align}
			W_{FKP}(\vx)\propto(C_\ell+N_\ell(\vx))^{-1},
			\label{eq:W_fkp}
		\end{align}
		where $C_\ell$ is the power spectra of the mode of interest and $N_\ell(\vx)=1/n_g(\vx)$ is the noise power spectra. This is the inverse variance weighting where $C_\ell$ accounts for the
		sample (cosmic) variance and $N_\ell$ accounts for variance contributed by the noise.
		Note that the FKP weights are sometimes written as, 
		$w\propto1/(1+n(\vx)C_\ell)$. This definition of weights is valid when the weights are applied to galaxies, while the weights we define in eq.~\eqref{eq:W_fkp} are applied to 
		the pixels of the over density maps.
		
		In the noise dominated regime, ($N_\ell\gg C_\ell$), the FKP weights reduce to $W_g$ in eq.~\eqref{eq:W_gF}. In general the FKP weights depend on the power spectra mode being 
		measured and for a tomographic survey with multiple redshift and $\ell$ bins,
		the proper use of FKP weight for optimal analysis can be computationally expensive. Hence pseudo-$C_\ell$ estimators with a single weighting scheme are usually preferred in practice. 
		For such a case, FKP like weights can be defined by fixing $C_\ell=C_0$, where $C_0$ is power spectra at some fixed chosen $\ell$. The overdensity field from 
		eq.~\eqref{eq:Pdelta_g}
		is then modified to
				\begin{align}
			\widehat{\delta}_{g,FKP_0}(\vx)=\frac{\mean{n_g(\vx)}}{\widebar{n}_g}\frac{1+\widebar{n}_g C_0}{1+\mean{n_g(\vx)}C_0}\delta_g(\vx),
			\label{eq:Pdelta_gFKP}
		\end{align}
		These weights are similar to those in eq.~\eqref{eq:W_gF}, except that $C_0$ here modulates the weights at the higher end to prevent few pixels from having very large weights 
		which can increase the cosmic variance.
				
		The maps for other observables such as galaxy shear can be defined analogously to eq.~\eqref{eq:Pdelta_g} as
		\begin{align}
			\widehat{\gamma}_{i,j}(\vx)=\frac{n_g(\vx)}{\widebar{n}_g}\gamma_{i,j}(\vx)=W_{\gamma}(\vx)\gamma_{i,j}(\vx),
			\label{eq:gammaW}
		\end{align}
		where the window $W_{\gamma}(\vx)$ depends on the observed number of galaxies in the pixel (as opposed the expected number of galaxies in $W_g$)
		\begin{align}
			W_{\gamma}(\vx)=\frac{n_g(\vx)}{\widebar{n}_g}=1+ F(C_i(\vx))+ \delta_g(\vx).
		\end{align}
		$W_{\gamma}(\vx)$ is relatively easier to estimate compared to $W_g$ since it is determined by the position of the source galaxies. That being said, the effects of 
		varying photometry on the shear estimation can be thought of as part of the window and hence		
		the problems associated with the window estimation we discuss later in this paper are applicable to galaxy shear estimation as well. The FKP like weights can also be defined for 
		shear noting that shape noise scales as $\sigma_e^2/n_g(\vx)$, resulting in 
		\begin{align}
			\widehat{\gamma}_{i,j,FKP}(\vx)=\frac{n_g(\vx)}{\widebar{n}_g}\frac{\sigma_e^2+\widebar{n}_g C_0}{\sigma_e^2+{n_g(\vx)}C_0}\gamma_{i,j}(\vx).
			\label{eq:gamma_fkp}
		\end{align}
		
		For the case of intensity mapping surveys, the window, $W(\vx)$, is same as the mean intensity term, $\widebar{I}$ (up to a constant), that is sometimes used 
		\citep[e.g.][]{Schaan2021}. In fact $W(\vx)$ is a generalization of the $\widebar{I}$ term in \cite{Schaan2021} as we allow for the effects such as foregrounds, detector 
		calibration, etc. to vary over the positions $\vx$. While \cite{Schaan2021} suggested that cross correlations may be useful in determining $\widebar{I}$ under certain 
		assumptions, unfortunately the cross correlations in a general case may not be of much help with the window modeling. 
		This is because different tracers/surveys have (at least partially) uncorrelated windows and hence the sensitivity of cross correlations to the window
		is very different from the sensitivity of the auto correlations. We will see an example of such an effect in section~\ref{sec:power_spectra}.
		
		In this work, we will restrict ourselves to the pseudo-$C_\ell$ like analysis and will use the weighting scheme of eq.~\eqref{eq:W_gF} and \eqref{eq:gammaW}, 
		which is more optimal in most cases than no weighting in \eqref{eq:delta_g} (see also discussion in section~\ref{ssec:shot_noise}). It should be remembered that 
		this weighting scheme increases the cosmic variance and in many practical applications 
		weighting of eq.~\eqref{eq:Pdelta_gFKP} and \eqref{eq:gamma_fkp} can be more optimal.

	\subsection{Shot noise}
		\label{ssec:shot_noise}
		Under the assumption that the sampling of the discreet tracers follow Poisson distribution, the noise in observed number of galaxies in each pixel is given by
		\begin{align}
			\mean{n_g^2(\vx)}=\mean{n_g(\vx)}.
		\end{align}
		When the galaxies are assigned weights, $w$, the noise changes to \citep[][ we assume weights are deterministic]{Bohm2014}
		\begin{align}
			\mean{n_{g,w}^2(\vx)}=\mean{n_g(\vx)}{w(\vx)^2}.
		\end{align}		
		
		The over density field of eq.~\eqref{eq:Pdelta_g} assigns weights of $1/\widebar{n}_g$ to each galaxy, in which case the noise is given by
		\begin{align}
			\mean{\delta_{N,W}^2}(\vx)=\frac{\mean {n_g(\vx)}}{\widebar{n}_g^2}={W_g(\vx)}\frac{1}{{\widebar{n}_g}}.
			\label{eq:N_g2}
		\end{align}
		Averaging over the whole survey, we get
		\begin{align}
			\widebar{\mean{\delta_{N,W}^2}}=\widebar{W}_g(\vx)\frac{1}{{\widebar{n}_g}}.
			\label{eq:PN_g}
		\end{align}
		
		Notice that in eq.~\eqref{eq:N_g2}, two powers of noise, $\delta_N^2$, depend on a single power of $W_g$. Therefore the noise is effectively multiplied by the window given by $\sqrt{W_g(\vx)}$
		\citep{FKP,Li2019}, 
		i.e., 
		\begin{align}
			\sqrt{\mean{{\delta_{N,W}^2}}}(\vx)=\sqrt{W_g(\vx)}\frac{1}{{\sqrt{\widebar{n}_g}}}.
			\label{eq:PN_g}
		\end{align}

		Similarly for the over density field in eq.~\eqref{eq:delta_g}, noise is given by
		\begin{align}
			{\mean{{\delta_{N}^2}}}(\vx)=\frac{1}{{\mean{{n}_g(\vx)}}}={\frac{1}{{W_g(\vx)}}}\frac{1}{{{\widebar{n}_g}}}.
		\end{align}
		Averaging over the whole survey, we get
		\begin{align}
			\widebar{\mean{{\delta_{N}^2}}}=\widebar {\left[\frac{1}{{W_g(\vx)}}\right]} \frac{1}{{{\widebar{n}_g}}}.
			\label{eq:N_g}
		\end{align}
		With $\widebar{W}_g(\vx)=1$ and $W_g(\vx)\in[0,\infty)$ by construction, it can be shown \referee{via Jensen's inequality that (see also appendix~\ref{app:noise_inverse_dist} for an alternate proof)}
		\begin{align}
			\widebar {\left[\frac{1}{{W_g(\vx)}}\right]}\geq \widebar{W}_g(\vx).
			\label{eq:noise_W_comp}
		\end{align}
		Thus the estimator in eq.~\eqref{eq:Pdelta_g} is in general has lower noise than the estimator of eq.~\eqref{eq:delta_g}. 
		Both estimator give very similar signal when $\widebar{W}_g(\vx)=1$.
		\refereetwo{Qualitatively, this is because the power spectra measurement, pseudo-$C_\ell$, depends on the window power spectra which is the second power of window while the noise depends 
		on the first power of window. Hence the choice of estimator has stronger effect on noise than on signal. We will also compute the response of the pseudo-$C_\ell$ power spectra estimator 
		to the window 
		power spectra in section~\ref{ssec:Dl_error} and show that the response is positive, implying that the estimator in eq.~\eqref{eq:Pdelta_g} will give larger signal in pseudo-$C_\ell$ 
		since it will in general have 
		higher window power. This further increases the signal to noise ratio of the estimator in eq.~\eqref{eq:Pdelta_g} with respect to the estimator in eq.~\eqref{eq:delta_g}.
		}
		
		The inequality in eq.~\eqref{eq:noise_W_comp} will get worse as the window becomes more complex ($W_g(\vx)$ distribution get wider). 
		Some times arguments are made in favor of using eq.~\eqref{eq:delta_g} to simplify the window 
		modeling but instead it becomes more important to use eq.~\eqref{eq:Pdelta_g} (or eq.~\eqref{eq:Pdelta_gFKP}) for more complex windows. 
		
		A corollary to eq.~\eqref{eq:noise_W_comp} is that with the estimator of eq.~\eqref{eq:delta_g}, 
		using $1/\widebar{n}_g$ to model the noise effects in the covariance (analytical or mocks with overly simplified window) will lead to under estimation of the such effects during the analysis, 
		unless the window effects are properly modeled. One simple method is to modify the effective number density of galaxies, but it can still leads to biases in the cross covariance terms of the form 
		$N_\ell C_\ell$, \referee{where $N_\ell$ is the noise power spectra}.
		
		Also, we only considered the galaxy shot noise as source of variance in this section. In appendix~\ref{appn:general_weighting}, we derive similar expression in presence of additional source of
		noise as well as weighting applied to galaxies. For the main part of this paper, we will continue using the galaxy shot noise only version of equations to keep the notation and discussion simpler. 
		The results of the paper do not change for the case considered in appendix~\ref{appn:general_weighting}.
		
		While in this paper we only consider noise correlated at zero lag, it is worth remembering that in the general case correlated noise will also depend upon the window and the 
		window for noise is 
		different from the window for the underlying over density field. Thus window effects on noise
		need to be modeled separately, both for mean noise subtraction from the power spectra measurement as well as in the covariance of both power spectra and correlation 
		functions. The 
		code, {\sc Skylens}, used in this paper is able to handle noise with any user input power spectra and window function.
		
	\subsection{Window bias}
		\label{ssec:map_err}
		In eq~\eqref{eq:W_gF}, we defined window as a function of some underlying contaminants. Even if those contaminants are known, we still need to estimate the response of the 
		selection 
		function to these contaminants in order to estimate the window. This is a non-trivial task and usually there will be some errors in estimating these responses, in which case 
		our 
		estimates of window can be biased.
		Without the loss of generality, we can write the estimated window as 
		\begin{align}
			\widehat{W}_g(\vx)=\left(1+m(\vx)\right)W_g(\vx),
			\label{eq:W_gm}
		\end{align}
		where $W_g(\vx)$ is the true underlying window and $m(\vx)$ denotes the relative error in the estimated window. 
		
		With the erroneous window defined in eq.~\eqref{eq:W_gm}, our estimate of the over-density fields become
		\begin{align}
			\widehat{\delta}_{g,W,m}(\vx)=&\frac{n_g(\vx)}{\widebar{n}_g}-\frac{\widehat{\mean{{n}_g(\vx)}}}{\widebar{n}_g},
		\end{align}
		where $\widehat{\mean{{n}_g(\vx)}}$ is the biased estimate of the expected number of galaxies. Writing the observed number of galaxies, $n_g(\vx)$ in terms of the true window 
		and
		the over density field, we get
		\begin{align}
			\widehat{\delta}_{g,W,m}(\vx)=&W_g(\vx)\delta_g(\vx)-\delta{W}_g(\vx)\nonumber \\ =&\frac{\widehat{W}_g(\vx)}{1+m(\vx)}\delta_g(\vx)-\delta{W}_g(\vx).
		\end{align}
		Where in the second step we wrote the true window in terms of the biased estimate, since that is the window we use in modeling. 
		When $m$ is small, the multiplicative bias can be estimated as $\sim -\widehat{W}_g(\vx)m(\vx)$.
		
		For the over density field defined in eq.~\eqref{eq:delta_g}, we can write the effects of bias as
		\begin{align}
			\widehat{\delta}_{g,m}(\vx)=\frac{\delta_g(\vx)-m(\vx)}{1+m(\vx)}. 
		\end{align}	
		Both estimators, $\widehat{\delta}_{g,W,m}$ and $\widehat{\delta}_{g}$ contain the 
		multiplicative bias of $O(m\delta)$ and an additive bias of $O({m})$.
		
		If $mW_g$ is a bias (as opposed to random noise) and is small, 
		we can assume the bias in the window is a linear combination of the underlying contaminants \citep[e.g][]{Leistedt2013}, $C_i$,
		\begin{align}
			\delta{W}_g(\vx)=m(\vx)W_g(\vx)\approx \sum\alpha_iC_i(\vx).
			\label{eq:map_dwindow}
		\end{align}
		Where $\alpha_i$ are the unknown amplitudes. It is important to stress here that we are only `assuming' the biases in the window function as linear function of contaminants 
		and 
		we do not assume the window function itself to be a linear function of the underlying contaminants. The additive biases can sometimes be constrained by cross-correlating the 
		over density maps with the maps of systematics \citep{Ross2012,Leistedt2013}. We will study the 
		effects of multiplicative bias in the section~\ref{ssec:Dl_error}.
		
		Finally it is worth mentioning that the multiplicative and additive biases in the shear estimation can also be thought of as part of the window in a similar vein as above and 
		our discussion of the window systematics here and in later sections is applicable to all tracers.
	
	\section{Power spectra}
		\label{sec:power_spectra}
		Once the over density maps are generated, the next step in the standard two point analysis is to compute the power spectra or the correlation functions of the maps. 
		In this section we will discuss the power spectra of the 
		windowed maps, the algorithm to model the window effects and the methods to account for the uncertainties in the window estimation.
		\subsection{Pseudo-$C_\ell$ power spectra}
			\label{ssec:pcl}
			Since the observed over density field is multiplied by the window, its Fourier transform\footnote{We will use Fourier transform to describe both flat sky Fourier transform and the 
			spherical harmonics transform in curved sky} is convolved with the Fourier transform of the window. Computing the power spectra of this 
			windowed field results in the pseudo-$C_\ell$ estimator ($D_\ell$), \refereetwo{whose expected value is related to} the true power spectra via a coupling matrix \citep{Hivon2002},
			\begin{equation}
				\widehat{D}_{\ell}=M_{\ell\ell'}C_{\ell'}+M_{\ell\ell'}^N N_{\ell'},
				\label{eq:Dl}
			\end{equation}
			where $M_{\ell\ell'}$ is the coupling matrix for the signal part and $M_{\ell\ell'}^N$ is the coupling matrix for the noise in the observed field (as discussed in 
			section~\ref{ssec:shot_noise} 
			noise and signal have a different window). \referee{Here summation over $\ell'$ is implied and throughout the paper we will use the Einstein summation convention to imply sum over 
			repeated indices.}
			To keep notation simpler, we will omit the the noise term from the equations. This will not affect our discussion since it is easy to 
			generalize to the noise case.
			
			The coupling matrices is given by 
			\begin{align}
            	M_{\ell,\ell'}={\frac{(2\ell'+1)}{4\pi}}\sum_{\ell''}
	             W_{\ell''}(2\ell''+1)
    	        &\wj{\ell}{\ell'}{\ell''}{s_1}{-s_1}{0}\nonumber \\ \times&\wj{\ell}{\ell'}{\ell''}{s_2}{-s_2}{0},
	        \label{eq:coupling_M}
    		\end{align}
			where \referee{where $\ell,\ell'$ are as defined in eq.~\eqref{eq:Dl}, $W_{\ell''}$ is the power spectra (pseudo-$C_\ell$) of the window defined at $\ell''$} 
			(cross power spectra of two windows for cross correlations), 
			$s_1,s_2$ are the spins of the two tracers being correlated to obtain the $\widehat{D}_{\ell}$.
			
			\refereetwo{ For the case of spin-2 quatities, e.g. galaxy shear or CMB polarization, the power spectra of $E$ and $B$ modes is given by, 
			\begin{align}
				D_{\ell}^{EE}=M^+ C_\ell^{EE}+M^- C_\ell^{BB}\label{eq:Dl_EE}\\
				D_{\ell}^{BB}=M^+ C_\ell^{BB}+M^- C_\ell^{EE}\label{eq:Dl_BB}.
			\end{align} 
			In the second equality we wrote E/B pseudo-power spectra in terms of true E/B power spectra using spin-2 coupling matrices, given by
			\begin{align}
            	M_{\ell,\ell'}^{\pm}=&{\frac{(2\ell'+1)}{4\pi}}\sum_{\ell''}
	            W_{\ell''}(2\ell''+1)\left(\frac{1\pm(-1)^{\ell+\ell'+\ell''}}{2}\right)
    	        \nonumber \\ \times&\wj{\ell}{\ell'}{\ell''}{2}{-2}{0}\wj{\ell}{\ell'}{\ell''}{2}{-2}{0}.
	        \label{eq:coupling_Mpm}
    		\end{align}
			In this section we will assume that the $B-$mode power spectra is zero, i.e. $C_\ell^{BB}=0$ and will therefore only use 
			$D_{\ell}^{EE}=M^+ C_\ell^{EE}$ and $D_{\ell}^{BB}=M^- C_\ell^{EE}$. We will discuss the more general case in section~\ref{ssec:EB_recons}. Hereafter, we will also drop the 
			superscripts $\pm, EE,BB$ on the $M$ and power spectra unless necessary for clarity.
			}
			
			In general $M_{\ell,\ell'}$ is a rectangular matrix with $\ell'_\text{max},\ell''_\text{max}\gg\ell_\text{max}$. In practice, the modern survey windows are large enough (narrow in 
			Fourier space) such that with $\ell_\text{max}\sim O(1000-5000)$ and $C_\ell\propto 
			\ell^{-2}$, $\ell'_\text{max}\sim\ell_\text{max}$ approximation is sufficient for accuracy of up to few percent near $\ell_\text{max}$ 
			(the examples shown later in this section satisfy this assumption). 
			With such an approximation, we need to estimate the window up to $\ell''\sim2\ell_\text{max}$. These approximation 
			however may not work if the window has large enough power out to very high $\ell''$ or in the case when the power spectra does not fall fast enough with $\ell$ (e.g. noise spectrum). In such a  
			case $\ell'_\text{max}\gg\ell_\text{max}$ maybe necessary or one may have to resort to trickery such as apodizing the window in order to tame the coupling matrix. 
			We will discuss an example of such effects in a later 
			section when we reconstruct power spectra from correlation functions.

		\subsection{{\sc Master} algorithm}
			As discussed in the previous section, one of the challenges in the pseudo-$C_\ell$ analysis is that $M_{\ell\ell'}$ is an O($\ell_\text{max}^2$) matrix where $\ell_\text{max}\sim1000-5000$ 
			or even larger. While computation of 
			$M_{\ell\ell'}$ can be easily handled on the modern day computers, it is still desirable to reduce the dimensionality of the $M_{\ell\ell'}$ for the purpose of sampling during the 
			inference.
			De-convolving unbinned $D_\ell$ to reconstruct $C_\ell$ by inverting eq.~\eqref{eq:Dl} is also not straight forward in the presence of large noise (remember noise has a different 
			coupling matrix and noise contributions scale as $M^{-1}[M^N(N_\ell)-\mean{M^N(N_\ell)}]$). Thus it is generally desirable to bin the noisy $D_\ell$ 
			measurements to reduce noise effects. 
			\cite{Hivon2002} presented the {\sc Master} algorithm which allows us to work with binned quantities, reducing the complexity of the problem to O$(N_b^2)$, where 
			$N_b$ is the number of bins.
			
			In the {\sc Master} algorithm, we simply write the eq.~\eqref{eq:Dl} in terms of the binned quantities,  
			\begin{equation}
				\widehat{D}_{\ell_b}=M_{\ell_b\ell_b'}C_{\ell_b'}+N_{\ell_b'}.
				\label{eq:Master}
			\end{equation}
			The binned $M_{\ell_b\ell_b'}$ is given by 
			\begin{equation}
				M_{\ell_b\ell_b'}=P_{\ell_b,\ell}M_{\ell\ell'}Q_{\ell',\ell_b'},
			\end{equation}
			where $P_{\ell_b,\ell}$ and $Q_{\ell',\ell_b'}$ are binning and inverse binning operations respectively and can be analytically written under the assumption 
			that $\ell(\ell+1)C_\ell\sim\text{constant}$, 
			\begin{align}
			P_{\ell_b,\ell}=\begin {cases} 
				\frac{1}{2\pi}\frac{\ell(\ell+1)}{\Delta \ell_b},  \ell\in b\\
				0 \text{ otherwise}
				\end{cases},
			\end{align}
			\begin{align}
			Q_{\ell,\ell_b}=\begin {cases} 
				\frac{2\pi}{\ell(\ell+1)},  \ell\in b\\
				0 \text{ otherwise}
				\end{cases},
			\end{align}
			where $\ell\in b$ is true when $\ell$ belongs to the bin centered on $\ell_b$ and $\Delta \ell_b$ is the bin size. In some implementations of the {\sc Master} algorithm,
			$C_{\ell}$ is assumed to be constant within the bin \citep[e.g.][]{Alonso2019}, in which case $P$ and $Q$ take values of 0 or 1. We will refer to such approximation 
			as {\sc cMaster}.
			
			The eq.~\eqref{eq:Master} can be inverted to reconstruct the power spectra from the pseudo-$C_\ell$
			\begin{equation}
				\widehat{C}_{\ell_b}=M_{\ell_b\ell_b'}^{-1}D_{\ell_b'},
				\label{eq:Master_inv}
			\end{equation}
			where we have taken the pseudo-inverse of the binned coupling matrix.
			
			\begin{figure}
					\centering
        	 		\includegraphics[width=\columnwidth]{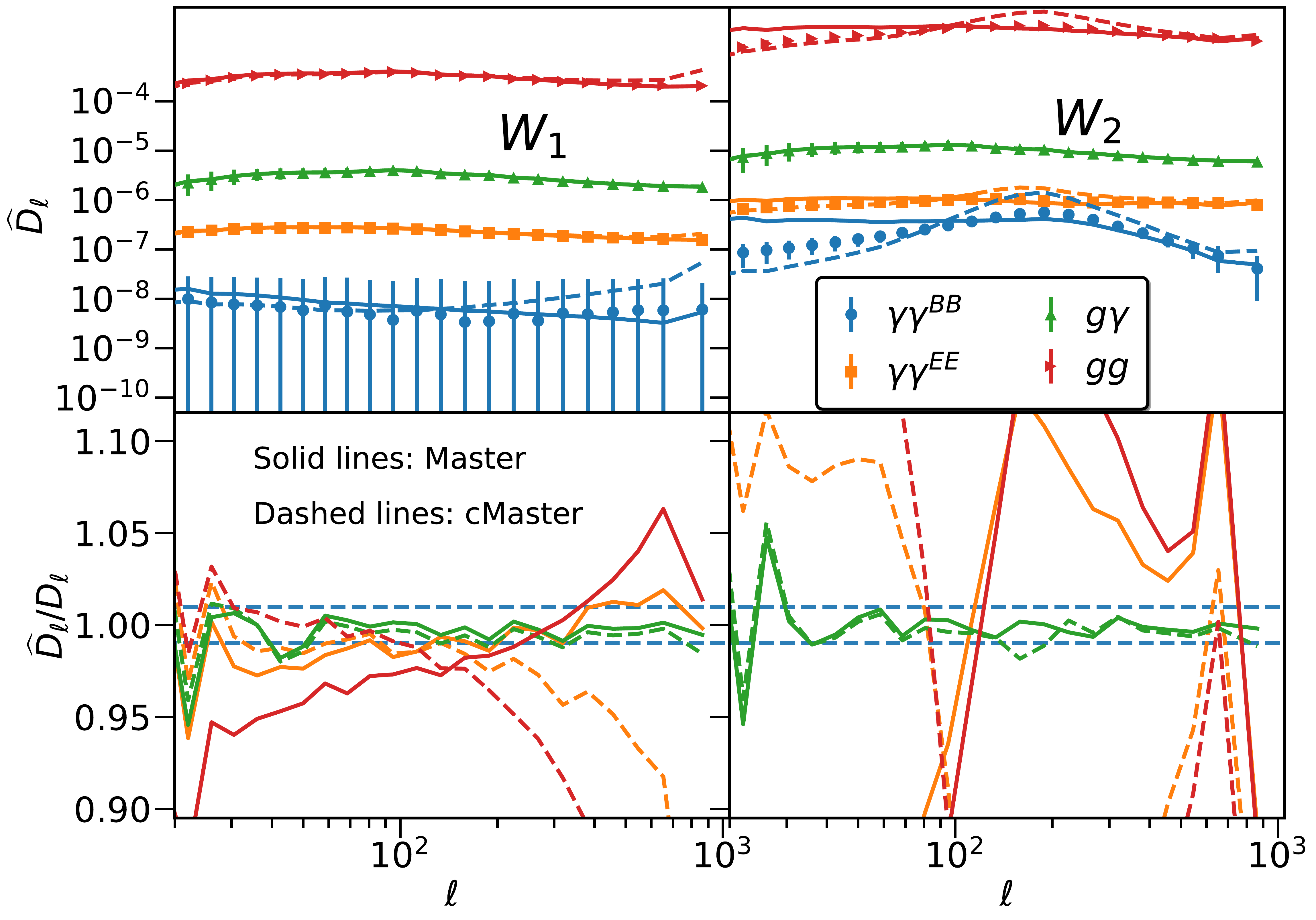}
					\caption{Comparison of binned pseudo-$C_\ell$ power spectra, $D_\ell$, obtained from 1000 gaussian simulations (solid points with errorbars) and predictions
					obtained using {\sc Master} \citep{Hivon2002}  and {\sc cMaster} \citep[e.g. ][]{Alonso2019} algorithms, for auto and cross correlations between shear and galaxies. 
					Left and Right panels show result from two different window functions, where $W_1$ is similar to the 
					expected galaxy shear window while $W_2$ is more complex. Lower panel on each side shows the ratio between the simulations and the predictions, with horizontal dashed lines 
					marking $\pm1\%$ bias regions. Both {\sc Master}  and {\sc cMaster} in general give biased results and the magnitude of the bias increases with the complexity of the window.
					}
					\label{fig:pcl_master}
			\end{figure}
			
			In fig.~\ref{fig:pcl_master}, we see the comparison of the $D_\ell$ predictions using both {\sc Master}  and {\sc cMaster} algorithms. \refereetwo{In fig.~\ref{fig:pcl_master} as well 
			as in most 
			other demonstrations later in the paper, we will imploy two choices of the window, $W_1$ and $W_2$. $W_1$ is chosen to be a more realistic and is similar to the galaxy shear window
			where the sampling is determined by galaxy positions and hence window power spectra is similar to galaxy power spectra. For $W_2$, we increase the complexity of window further by 
			raising the power spectra of window further at some scales (see appendix~\ref{app:sims} for further details). $W_2$ acts as a stress tests for our methodology and comparison with 
			$W_1$ also allows us to see where and how the biases are introduced by different methods.} 
			
			\refereetwo{In fig.~\ref{fig:pcl_master} we  observe that as the complexity of the window increases,
			both algorithms result in biased predictions. This is because coupling matrix becomes broader with more complex window and since $D_{\ell_b}$ is effectively a weighted sum 
			of $C_\ell$, a broader coupling matrix increases the impact of biases introduced by the 
			approximations made in the definition of binning operators $P$ and $Q$. } Furthermore, {\sc cMaster} has larger biases since it makes worse approximations in  $P$ and $Q$. 
			\referee{ These biases also propagate to the deconvolved power spectra (see fig.~\ref{fig:master_comp}) and
			necessitate that some corrections be applied to the theory power spectrum before it can be compared with the biased estimators of either $D_\ell$ or denconvolved power spectra 
			\citep[see discussion in ][]{Alonso2019}, i.e. we define:
			\begin{equation}
				{C}_{\text{cMaster},\ell_b}=M_{\text{cMaster},\ell_b,\ell_b'}^{-1}B_DM_{\ell' \ell''}C_{\ell''}=\mathcal M_{\text{cMaster},\ell_b,\ell''}C_{\ell''}.
				\label{eq:cMaster_th}
			\end{equation}
			\refereetwo{$B_D$ is the binning operator similar to $P$ and is defined in next section. In the second equality we defined $\mathcal M_{\text{cMaster},\ell_b,\ell''}=M_{\text{cMaster},\ell_b,\ell_b'}^{-1}B_DM_{\ell' \ell''}$}.
			Using such an estimator is not strictly necessary, as one can simply work with the binned pseudo power spectra instead, i.e. 
			\begin{equation}
				{D}_{\ell_b}=M_{\ell_b\ell'}C_{\ell'}, 
				\label{eq:Dlb_th}
			\end{equation}
			where $M_{\ell_b\ell'}= B_D M_{\ell\ell'}$. This is a simple convolution+binning operation 
			and will not be branded as {\sc Master} algorithm in this paper.
			}
			
			In the next section
			we discuss the proper expression for $P$ and $Q$ which lead to recovery of unbiased results which can be compared directly with theory power spectra.
			
		\subsection{{\sc $i$Master}}
			\label{ssec:iMaster_cl}
			As discussed in the previous section, 
			the biases in the {\sc Master} algorithm are sourced by the assumptions made in implementing 
			the effects of  binning. Now we derive the {\sc Master} algorithm without making such assumptions.
			
			\refereetwo{The binned version of eq.~\eqref{eq:Dl} can be written as 
			\begin{align}
				B_{D,\ell_b,\ell} D_{\ell}=B_{D,\ell_b,\ell} M_{\ell\ell'}B_{C,\ell',\ell_b'}^{-1}B_{C,\ell_b',\ell'}C_{\ell'}+\nonumber\\
											B_{D,\ell_b,\ell} M_{\ell\ell'}^NB_{N,\ell',\ell_b'}^{-1}B_{N,\ell_b',\ell'} N_{\ell'},
											\label{eq:Dl2}
			\end{align}	}	
			where $B$ is the binning operator. We introduced separate binning \refereetwo{operators, $B_C$, $B_D$ and $B_N$, which operate on $C_\ell'$, $D_\ell$, $N_\ell'$}
			since the coupling matrix is not symmetric in general and $\ell$, $\ell'$ can 
			have different range. For examples shown throughout this work we will make the approximation that $\ell_\text{max}\sim \ell'_\text{max}$, (coupling matrix is not too wide) in which 
			case $B_D=B_C$ \refereetwo{can be used. We will simply assume $B_D=B_C=B$ and omit the subscripts hereafter unless required for clarity. 
			Furthermore we assume noise is estimated and subtracted before binning and will neglect the effects of $B_N$ as well. We will not assume any particular form for the binning operator 
			and will keep our discussion general so that our method does not depend on the binning operator unlike the standard {\sc Master} algorithm.
			We will discuss some particular forms of binning operator 
			later in this section and the form of the binning operator used in examples of this paper is defined in eq.~\eqref{eq:bin_default}.
			}
			
			Binning data in general leads to loss of information and hence the binning operation can not 
			be inverted. Thus $B^{-1}$ in general is not defined. However, if we have good model for the underlying signal, then it is possible to obtain $B^{-1}$.
			In \cite{Hivon2002}, model for $C_\ell$ was assumed to be $\ell(\ell+1)C_\ell\sim\text{constant}$ which allowed for an analytical expression for $Q_{\ell,\ell_b}$ which is 
			the inverse of binning operator $P_{\ell_b,\ell}$. However, since we hopefully have a better model for power spectra, the assumptions about $C_\ell$ are not 
			necessary and we can simply write the inverse of binning operator as
			\begin{equation}
				B^{-1}_{C,\ell,\ell_b}=\begin {cases} 
				\frac{C_{\ell}}{C_{\ell_b}},  \ell\in b\\
				0 \text{ otherwise}
				\end{cases},
				\label{eq:B_inv}
			\end{equation}
			where $C_{\ell_b}=BC_\ell$ is obtained after binning the model $C_\ell$ and can be described as power spectra at some effective center of the bin, $\ell_b$. 
			The proper choice of $\ell_b$ is the effective $\ell$ at which the binned $C_\ell$ is measured and is given by
			\refereetwo{
			\begin{equation}
				\ell_b=\frac{B_{C,\ell_b,\ell} \ell C_\ell}{B_{C,\ell_b,\ell} C_\ell}.
				\label{eq:ell_b}
			\end{equation}}
			
			\refereetwo{
			The {\sc $i$Master} binned coupling matrix is then
			\begin{equation}
				M_{\ell_b\ell_b'}=B_{D,\ell_b,\ell} M_{\ell\ell'}B_{C,\ell',\ell_b'}^{-1}.
				\label{eq:coupling_M_binned}
			\end{equation}}
			
		\begin{figure*}
				\begin{subfigure}[t]{\columnwidth}
    	    		\centering
        	 		\includegraphics[width=\columnwidth]{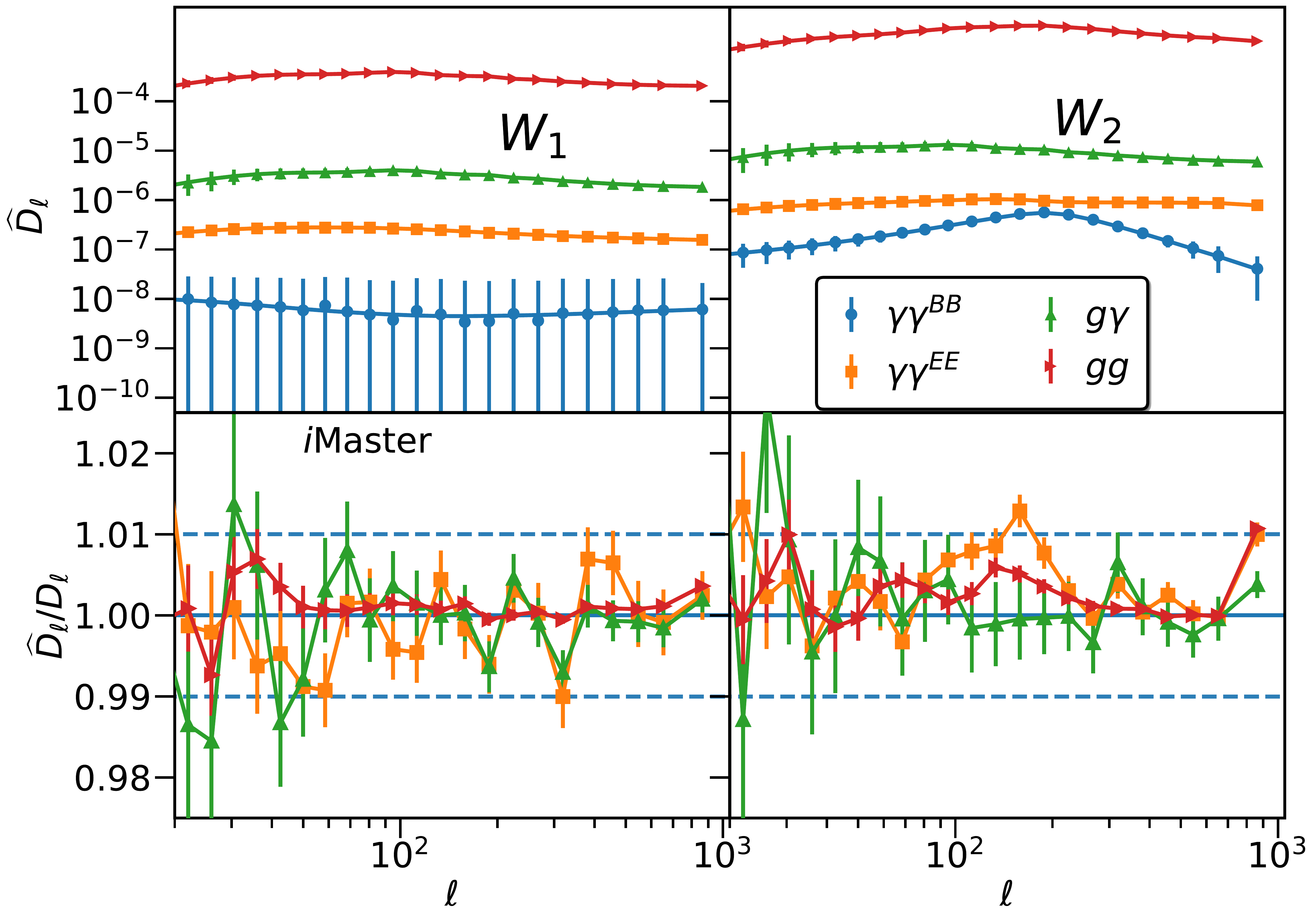}
					\caption{Pseudo $C_\ell$}
					\label{fig:pcl_imaster}
				\end{subfigure}
				\begin{subfigure}[t]{\columnwidth}
	         		\includegraphics[width=\columnwidth]{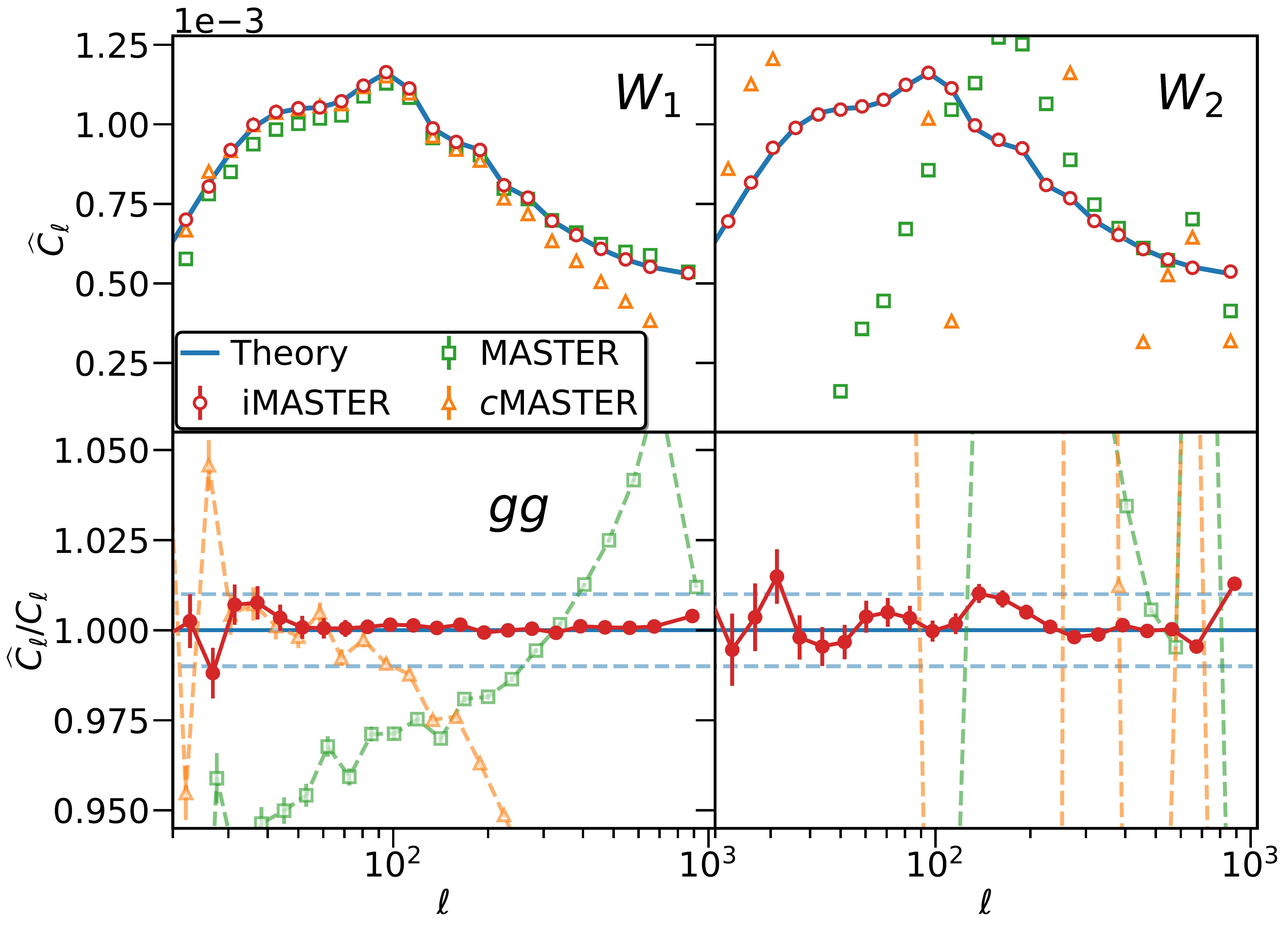}
					\caption{Galaxy-galaxy, $C_\ell$
					}
					\label{fig:gg_imaster}
				\end{subfigure}
				\begin{subfigure}[t]{\columnwidth}
	         		\includegraphics[width=\columnwidth]{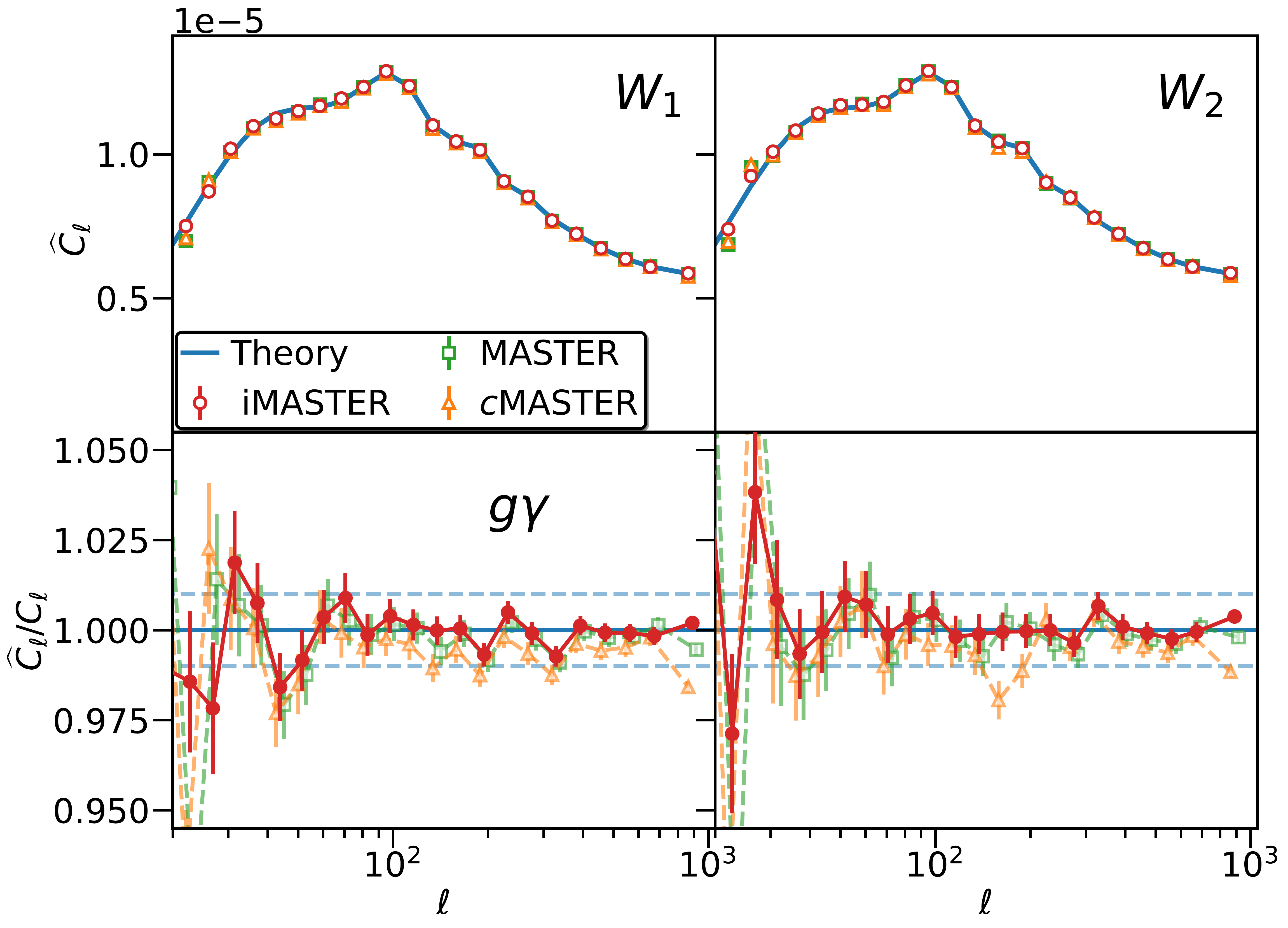}
					\caption{Galaxy-shear, $C_\ell$
					}
					\label{fig:ggl_imaster}
				\end{subfigure}
				\begin{subfigure}[t]{\columnwidth}
	         		\includegraphics[width=\columnwidth]{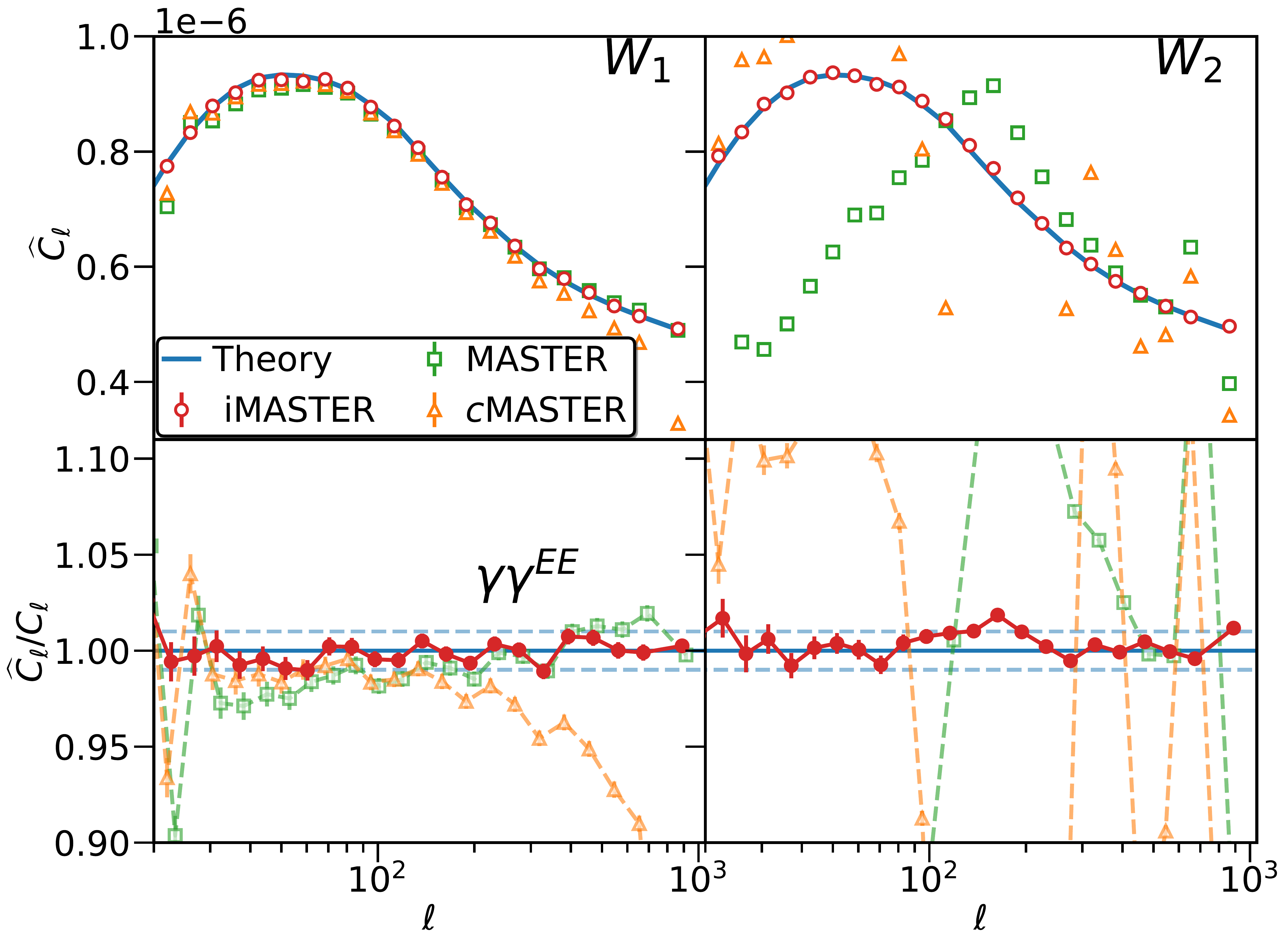}
					\caption{Shear-shear, $C_\ell$
					}
					\label{fig:ll_imaster}
				\end{subfigure}

				\caption{Comparisons of $D_\ell$ and $C_\ell$ of different observables, obtained using two different window functions, $W_1$ and $W_2$ (see appendix~\ref{app:sims}). 
				In the upper panel of each figure, the solid lines show the model predictions while the points represent the mean and the error on the mean from 1000 gaussian simulations. The lower 
				panel show the ratio of simulations to the model predictions. a) 
				Pseudo-$C_\ell$ power spectra, $D_\ell$, for auto and cross correlations of galaxy positions and galaxy shear obtained using {\sc $i$Master} algorithm. Model predictions agree with the 
				simulations to better 1\% after subtracting the noise.
				b-d) $C_\ell$ obtained by deconvolving power spectra using standard {\sc Master} algorithm, {\sc $i$Master} algorithm and the {\sc cMaster} algorithm.  
				{\sc $i$Master} gives unbiased power spectra (within 1\% error) while {\sc Master} and {\sc cMaster} algorithm gives biased results on scales where window has large power. The 
				biases become 
				worse as the complexity of the window increases. In $g\gamma$, where the window power spectra is small (window of galaxies and shear are uncorrelated) the biases in {\sc Master}
				and {\sc cMaster} are small. The biases are largest in auto correlation with $W_2$ window which has large power out to high $\ell$ (see appendix~\ref{app:sims}).
				\referee{As mentioned in \citet{Alonso2019}, the {\sc cMaster} like results require convolving the theory power spectra with a coupling matrix in order to perform an unbiased
				comparison between theory and data (see also eq.~\eqref{eq:cMaster_th}). 
				The {\sc $i$Master} requires no such corrections and the measurements can directly be compared with underlying binned theoretical model or the 
				equivalent unbinned power spectra computed only at the effective scale $\ell_b$ of each bin.}
  					}	    	     
					\label{fig:master_comp}
		     \end{figure*}
		     
		  	Fig.~\ref{fig:pcl_imaster} shows the comparison of $D_\ell$ obtained using the {\sc $i$Master} algorithm with that of 1000 gaussian simulations, similar to fig.~\ref{fig:pcl_master}. 
			The model predictions are consistent with those of simulations to well within $1\%$, even for the case of more complex window $W_2$. Figs.~\ref{fig:gg_imaster}-\ref{fig:ll_imaster}
			show the comparison of the $C_\ell$ reconstructed from the $D_\ell$ of simulations by convolving with the (pseudo) inverse of binned coupling matrix as described in 
			eq.~\eqref{eq:Master_inv}. Here again we observe that the {\sc $i$Master} yield unbiased results to within $1\%$ while both {\sc Master} and  {\sc cMaster}  yield biased results as in 
			the case of $D_\ell$ predictions in fig.~\ref{fig:pcl_master}.
			
			\referee{It is important to stress here that the $C_\ell$ from the {\sc $i$Master} algorithm can be directly compared with the unbinned theory computed at $\ell_b$, i.e. the binning 
			is performed analytically. This algorithm also does not 
			require any additional correction to be applied unlike the method in eq.~\eqref{eq:cMaster_th} \citep{Alonso2019}. \refereetwo{Furthermore, since the algorithm allows for 
			cleaner reconstruction of $C_\ell$, it is also more optimal in the analysis with scale cuts as it prevents the mixing of information between different scales in a much cleaner way. 
			For example, in eq.~\eqref{eq:Dl2} if we have scale cuts 
			of form of $\ell'<\ell_\text{max}'$, {\sc $i$Master} prevents the information from  $\ell'>\ell_\text{max}'$ modes leaking into our analysis. At the same time the algorithm also
			captures the information about $\ell'<\ell_\text{max}'$ modes which had leaked into high $\ell$ modes ($\ell>\ell_\text{max}'$) of the pseudo-$C_\ell$ estimator ( this requires $\ell_\text{max}
			>\ell_\text{max}'$
			in eq.~\eqref{eq:Dl2} and also using different binning operators $B_D$ and $B_C$).}
			These are the primary advantages of the {\sc $i$Master} algorithm.}
			
			\refereetwo{Due to $B^{-1}$, the estimator is now dependent on the underlying theoretical model. If one feels uncomfortable with such dependence, a half binned version, 
			$M_{\ell_b\ell'}=B_D M_{\ell\ell'}$ (see eq.~\eqref{eq:Dlb_th}) can be used, with larger computational costs (both time and system memory). 
			As discussed earlier, {\sc $i$Master} is also more optimal in terms of scale cuts and preventing mixing of information between different scales. 
			It may be advantageous to use {\sc $i$Master} in the form similar to eq.~\eqref{eq:cMaster_th}, i.e., 
			\begin{equation}
				{C}_{\text{\sc $i$Master},\ell_b}=M_{\text{$i$Master},\ell_b,\ell_b'}^{-1}B_DM_{\ell' \ell''}C_{\ell''}=\mathcal M_{\text{$i$Master},\ell_b,\ell''}C_{\ell''}.
				\label{eq:iMaster_fixed}
			\end{equation}
			where $\mathcal M_\text{\sc $i$Master}$ is computed at fiducial cosmology and is kept fixed.
			To reiterate, one of the advantages of {\sc $i$Master} is that ${C}_{\text{\sc $i$Master},\ell_b}={C}_{\ell_b}$ and we can use the un-binned theory to compare with the measurements. 
			Only in the case where the model varies strongly enough such that variations in {\sc $i$Master} are significant compared to noise (while such variations are found to be small in our 
			examples, similar tests should be performed at the time of analysis), 
			using the form of eq.~\eqref{eq:iMaster_fixed} will be necessary, where theory is properly binned using a fixed $\mathcal M_\text{\sc $i$Master}$ computed with fiducial model.
			It should also be noted that the model dependence of {\sc $i$Master} is weak, as $B^{-1}$ is only sensitive to the 
			slope of the power spectra within the bin. As long as the ratio does not vary significantly the model dependence should not introduce any significant biases. In 
			fig.~\ref{fig:master_comp}, the $\chi^{2}$ per bin for different curves obtained using {\sc $i$Master} is always less than $2/25$ and for noiseless simulations (not shown) is 
			less than $0.5/25$. Therefore the model dependence introduced by {\sc $i$Master} is not much different from the model dependence from other components in the analysis, e.g. converting 
			distances to redshifts or computing covariance matrices at fixed cosmology.
			If the model dependence of {\sc $i$Master} does matter in the inference problem, i.e.
			the bias due to assumptions in $B^{-1}$ is larger than the uncertainties, 
			it is also likely to be an indicative of a problem with binning, namely that bins are too wide and it maybe better to move to narrower bins to capture more information. 
			It is not optimal to use overly broad bins with {\sc $i$Master}. If using narrow bins is not possible, then obtaining model using eq.~\eqref{eq:iMaster_fixed} may be necessary.} 
	
			\refereetwo{Now we discuss some particular forms of the binning operators. It should be remembered that in deriving the {\sc$i$Master} algorithm we did not assume any particular form 
			of the binning operators and any sensible choice of binning will work.}
			
			For noisy measurements with a given covariance, the binning operator can be defined using the optimal estimator of the mean $D_\ell$ within the bin (assuming gaussian distribution),
			\begin{equation}
				B=\frac{U^T{\text{Cov}}^{-1}}{U^T\text{Cov}^{-1}U},
				\label{eq:B_opt}
			\end{equation}
			where Cov is the covariance of the unbinned $D_\ell$. $U$ is $n_{bins}\times n_\ell$ matrix given by
			\begin{equation}
				U_{\ell_b,\ell}=\begin {cases}1,  \ell\in b\\ 0 \text{ otherwise}.
				\end{cases}
			\end{equation}
			The optimal binning operator in eq.~\eqref{eq:B_opt} depends on the unbinned covariance of the power spectra. While estimation of this covariance is expensive, it is naturally
			obtained in the intermediate steps when computing the analytical covariances of binned power spectra. The binning operator can be obtained at the same time as computing 
			covariance
			with little additional computing cost. $B$ also depends on the particular power spectra being considered and hence each cross and auto correlation in a tomographic analysis will 
			require different $B$. If such properties of $B$ are not desirable, an approximation to B can be used where each $\ell$ is weighted by the effective number of modes 
			(ignoring the effects of noise)
			\begin{equation}
				B_{\ell_b,\ell}\propto \begin {cases} \frac{2\ell+1}{(2\ell_b+1)\Delta \ell},  \ell\in b\\ 0 \text{ otherwise}. 
				\end{cases}
				\label{eq:bin_default}
			\end{equation}	
			Simulations and theory calculations shown in this work use the binning operators from eq.~\eqref{eq:bin_default}.

		\subsection{Window bias}
		\label{ssec:Dl_error}
			In this section we discuss the response of the pseudo-$C_\ell$ estimator to the errors in estimating the window functions. We begin by noticing that the $D_\ell$ is symmetric 
			in response to the window and the underlying over density field and therefore we can write (see appendix~\ref{app:Ang_PS}):
			\begin{equation}
				D_\ell=M^W_{\ell,\ell''}W_{\ell''},
			\end{equation}
			where $M^W_{\ell,\ell''}$ is the response of $D_\ell$ to the window and depends on the $C_\ell$. Relative error in $D_\ell$ due to errors in window power spectra can be 
			written as 
			\begin{equation}
				\frac{\partial\log D_\ell}{\partial W_{\ell''}}=\frac{1}{D_\ell}M^W_{\ell,\ell''}.
				\label{eq:Dl_Rerror}
			\end{equation}
			$\frac{1}{D_\ell}M^W_{\ell,\ell''}$ can be used to define the upper limits on the bias in window power spectra $\delta W_{\ell''}$ given an upper limit on the
			percentage bias $D_\ell$ that is acceptable.
			
			Given the multiplicative and additive errors in the window as defined in section~\ref{ssec:map_err}, we can also write the bias in $D_\ell$ as
			\begin{equation}
				\delta D_\ell=M^W_{\ell,\ell''}m_{\ell''}+ a_\ell,
				\label{eq:Dl_error}
			\end{equation}
			where $m_{\ell''}$ is the power spectra of the multiplicative bias ($m(\vx)W(\vx)$) and $a_\ell$ is the power spectra of the additive bias. 
			It is important to note that multiplicative and additive biases do not necessarily have to be present together. For example, from eq.~\eqref{eq:W_gm}, 
			it is possible that the estimated and true windows have the very similar power spectra but the two window are not very correlated. In such a case, 
			window errors will behave like noise, resulting in additive bias $a_\ell$ but negligible multiplicative bias. Efficacy of window weighting will also be 
			lost in such a case.
			
			For window biases which may be described by eq.~\eqref{eq:map_dwindow}, under the assumption that the \referee{contaminant} maps are not correlated, we can write 
			\citep{Ross2012,Leistedt2013,Elsner2017}
			\begin{equation}
				a_\ell=\sum_i\alpha_i^2 D_{i,\ell},
				\label{eq:Dl_a_error}
			\end{equation}
			$D_{i,\ell}$ is the power spectra of the \referee{contaminant} maps.
			$\alpha_i$ can be estimated by cross correlating the estimated over density map (which includes systematics) with \referee{contaminant} map as
			\begin{align}
				\alpha_i&=\frac{D_{i,\ell}\text{Cov}^{-1}\mean{\widehat{\delta}_{g,W}C_{i}}_\ell}{D_{i,\ell}\text{Cov}^{-1}D_{i,\ell}},
				\label{eq:alphai}\\
				Var(\alpha_i)&=\frac{1}{D_{i,\ell}\text{Cov}^{-1}D_{i,\ell}} \label{eq:var_alphai}.
			\end{align}
			$\mean{\delta_{g,W}C_i}_{\ell}$ is the cross power spectra between the over density map and the contaminant map. 
			$\text{Cov}$ is the joint covariance of the auto and cross correlations and eq.~\eqref{eq:var_alphai} denotes the variance of $\alpha_i$.
			The method is only valid under the assumptions of linear expansion in eq.~\eqref{eq:map_dwindow} and that the \referee{contaminant} maps are independent. 
			If the \referee{contaminant} maps are not independent, one can use a 
			PCA like method to project the \referee{contaminant} maps into maps of orthogonal linear combinations which are uncorrelated and then define $\alpha_i$ and cross correlations over the 
			new set of maps.
			
			The cross correlation in eq.~\eqref{eq:alphai} is not sensitive to the $m_\ell$ as it shows up in the third order term, whose expectation is zero when window and the 
			underlying overdensity field are uncorrelated. If the corrections from the above procedure are reliable and are applied to the window, it may reduce the 
			effects of multiplicative bias as well.
			
\begin{figure*}
			\begin{subfigure}[t]{\columnwidth}
    	    		\centering
        	 		\includegraphics[width=\columnwidth]{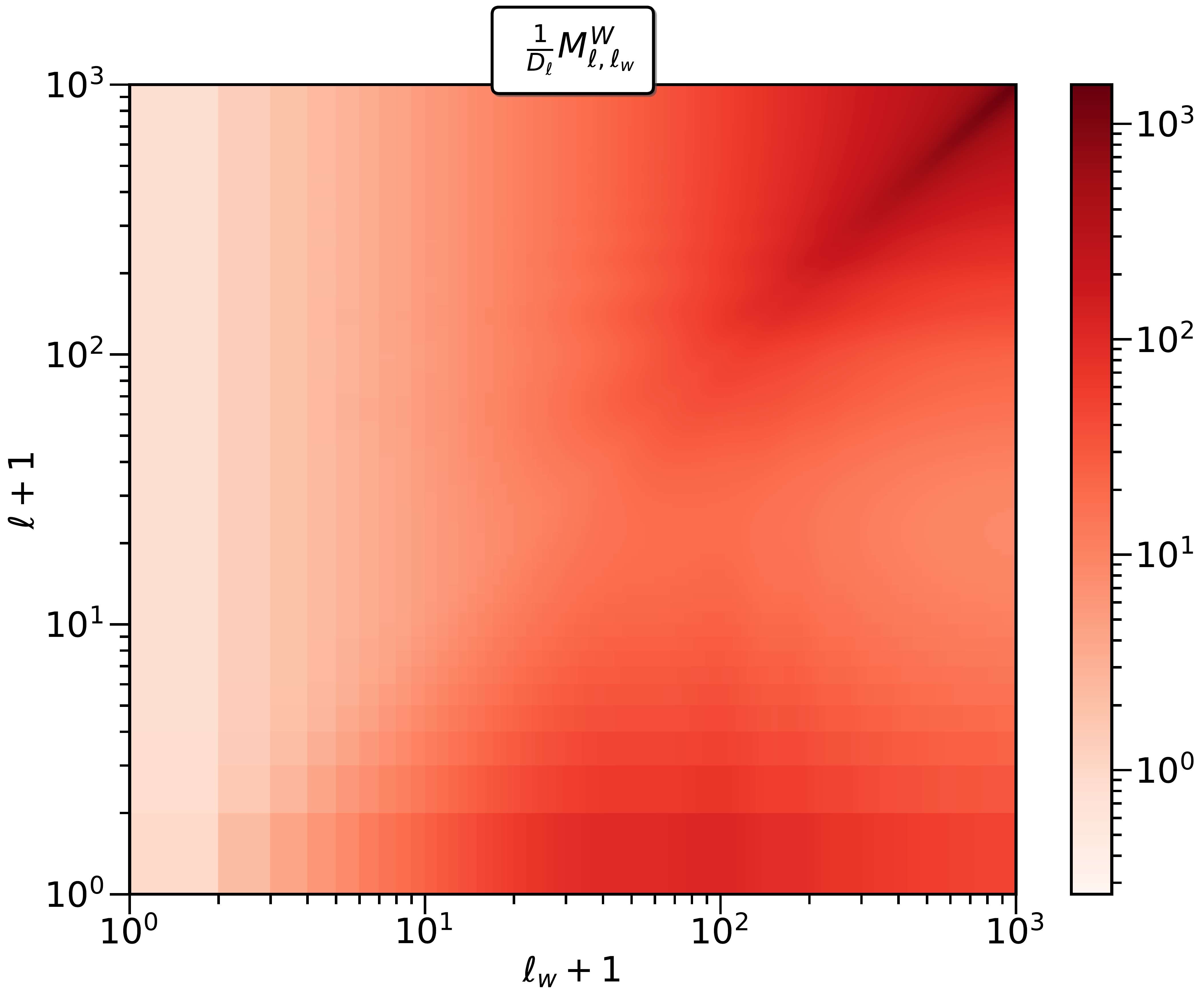}
					\caption{}
					\label{fig:window_M}
			\end{subfigure}
			\begin{subfigure}[t]{\columnwidth}
    	    		\centering
        	 		\includegraphics[width=\columnwidth]{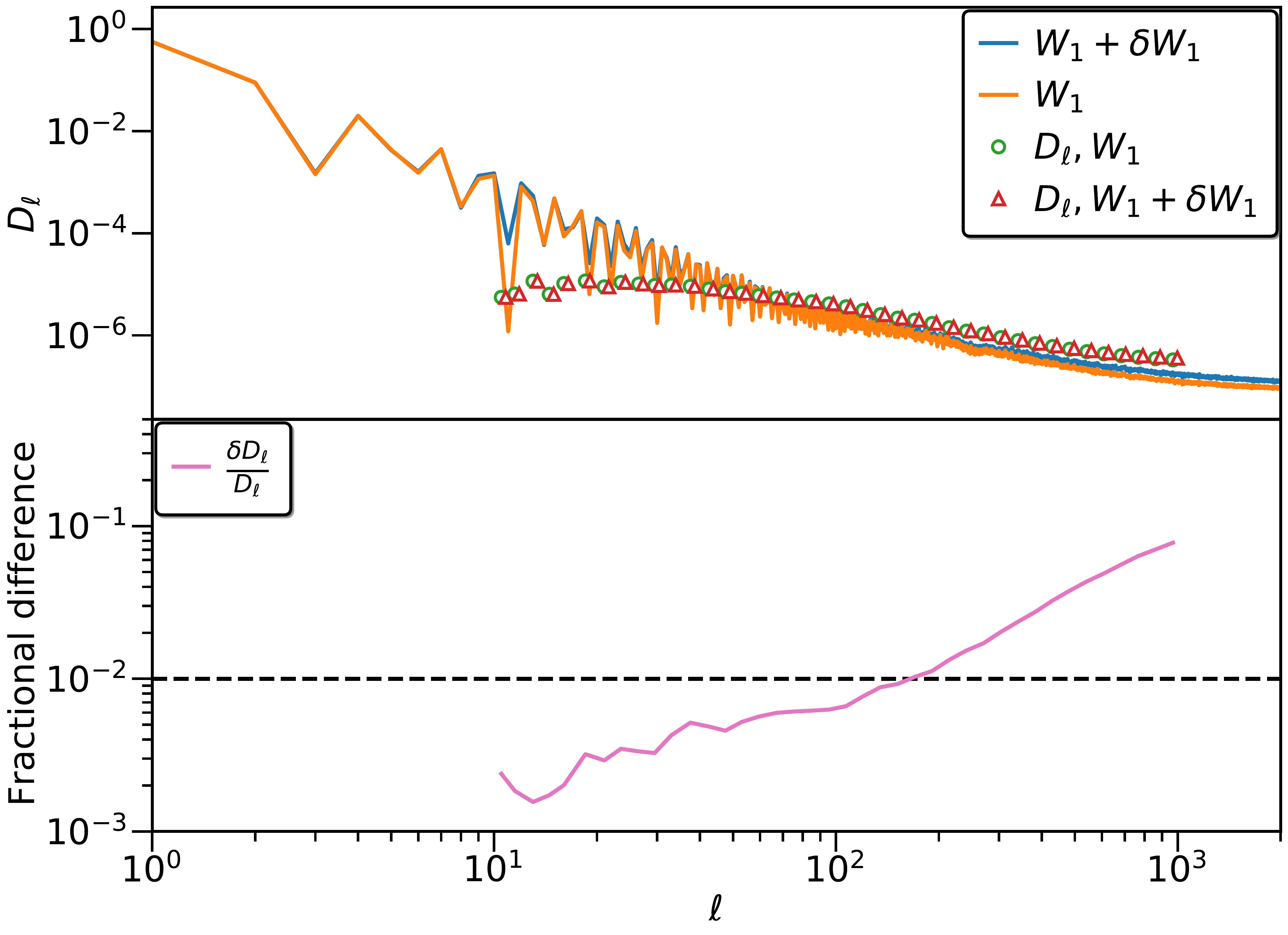}
					\caption{}
					\label{fig:window_dl}
			\end{subfigure}
			\caption{ a)  The matrix defined in eq.~\eqref{eq:Dl_Rerror}, which relates the errors in the window power spectra, $\delta W_\ell$ to the relative errors in the 
			$D_\ell$, i.e. $\frac{\delta D_\ell}{D_\ell}=\left(\frac{1}{D_\ell}M^W\right) \delta W_\ell$. Note that $\ell$ values
			are shifted by 1 to show the $\ell=0$ column which relates the $f_\text{sky}$ error to errors in $D_\ell$. At high $\ell$ the matrix has strong values around the diagonal
			which increases the relative sensitivity of $D_\ell$ to the window power spectra. b) Demonstration of the multiplicative bias introduced in $D_\ell$ by errors in $W_\ell$ (there is no 			additive 
			bias in this figure). \referee{In the upper panel, solid orange line shows the default window while blue line shows the perturbed window. Green circles and red triangles show the 
			$D_\ell$ obtained using these windows and the lower panel shows the fractional bias introduced in $D_\ell$.}
			Here the window was perturbed by simply adding a 
			white noise with $N_\ell\sim W_{\ell=500}$ power spectra ($N_{\ell<10}=0$ and $\delta f_\text{sky}=0$ by construction). 
			$W_1$ refers to the realistic galaxy shear like 
			window (see appendix~\ref{app:sims}).
			}
			\label{fig:window_error}
		\end{figure*}

			For the case of multiplicative bias, fig.~\ref{fig:window_M} shows the $\frac{1}{D_\ell}M^W_{\ell,\ell''}$ matrix. 
			At low $\ell$ the matrix has nearly scale independent effects along the column, especially 
			at $\ell=0$. This accounts for the
			effects of the misestimation of the largest scales of the window, primarily sourced by the mask, and includes the effects of the mean of window or the $f_\text{sky}$ factor. 
			The diagonal of the matrix is sub-dominant at low $\ell$ but it increases as $\sim2\ell+1$,
			becoming large at high $\ell$. Typically when modeling the window in power spectra such high $\ell$ effects are ignored. However, in fig.~\ref{fig:window_dl}, we see that 
			this effect leads to rapidly increasing biases at higher $\ell$, which dominate the information during the cosmological inference. Thus it is important to carefully model window
			out to high $\ell$ and properly account for any uncertainties in such modeling. Current modeling, e.g. \cite{Ross2012,Ross2020}, uses {\sc Healpix} maps with resolution of 
			order $N_\text{side}=512$, which may not be enough in the case of moderately complex windows even for $\ell_\text{max}\sim1000$ (roughly similar to 
			scale cuts in current weak lensing $3\times2$ analysis) and in the upcoming analysis such with 
			LSST where we may wish to extend the analysis to $\ell_\text{max}\gtrsim5000$, a much more careful characterization of the window will be needed.
			
			To account for uncertainties in window effects, one can use some parametric function for $m_\ell$ which can be added to the $D_\ell$ model using eq.~\eqref{eq:Dl_error} 
			and its parameters can be marginalized over. We can also use the analytical marginalization \citep{Bridle2002}, in which case the 
			the contributions from window uncertainties can be added to the covariance of the $D_\ell$, and the additional contribution to the covariance can be written as
			\begin{equation}
				Cov^W(D_\ell,D_{\ell'})\approx M^W m_\ell m_{\ell}^T M^{W,T}+a_\ell a_{\ell}^T,
			\end{equation}
			where $a_\ell$ and its uncertainty can be obtained from eqs.~\eqref{eq:Dl_a_error}--\eqref{eq:var_alphai}.
			In general, if we have an estimation of the covariance of the window power spectra, $W_\ell$, we can write
			\begin{equation}
				Cov^W(D_\ell,D_{\ell'})\approx M^W Cov (W_\ell,W_{\ell'}) M^{W,T}+Cov (W_\ell,W_{\ell'}).
			\end{equation}
			This expression is an approximation since a full covariance with the window and the over density field will require an expansion of 8-point function, which is tedious and 
			computationally expensive  
			even for the case of gaussian field. A detailed study of the window uncertainties in a realistic setting of LSS windows for the case of DESI survey will be 
			presented in an upcoming work (Karim, Rezai \& Singh in prep).
			
			There are also cases where the window is well estimated but the multiplicative error can still have significant contributions to the covariance. One such case is when 
			estimating the jackknife covariance, where each jackknife sample has a slightly different window and these seemingly minor differences can bias the 
			covariance (Yu, Singh et al. in prep).
			
			Finally, as discussed in section~\ref{sec:maps}, it is also tempting to conclude that we can use cross correlations 
			between different tracer/surveys to `self calibrate' the effects of window. 
			However, this in general does not work unless the windows between tracer/surveys are perfectly correlated. An example of such an effect can be observed on 
			figs.~\ref{fig:pcl_master} and 
			~\ref{fig:master_comp}, where galaxies and shear have uncorrelated window by design and hence the cross correlation, $g\gamma$, is almost completely free of the window 
			effects except for the very large scales (low $\ell$), where windows are partially correlated due to the mask effects. Thus it is in general not possible to use the cross 
			correlation, $g\gamma$, 
			to understand the window effects in auto correlations, $gg$ and $\gamma\gamma$ in our case.
			
	\section{Correlation functions}
		\label{sec:correlation_function}
		In this section we discuss the two point function measurements in the configuration space, namely the correlation functions. We begin with brief introduction of correlation functions 
		and will discuss the estimators, effects of windows, the methods to apply scale cuts and the 
		generalization of the {\sc $i$Master} algorithm to reconstruct the power spectra from the correlation functions.

			The curved sky correlation function can be written as the Hankel transform of the power spectra,
			\begin{equation} 
				\xi(\theta)=\sum_{\ell} \frac{2\ell+1}{4\pi} {_{s_1}}d_{\ell,s_2}(\theta) C_\ell =\mathcal H_{\theta,\ell} C_{\ell},
				\label{eq:xi0}
			\end{equation}
			where $_{s_1}d_{\ell,s_2}$ is the wigner-d matrix, with $s_1,s_2$ being the spins of tracers being correlated and in the second equality we have written the sum as a  matrix 
			multiplication, with $\mathcal H$ being the Hankel transform operator. The sum is over all $\ell$ and therefore 
			computing $\xi$ to an arbitrary accuracy can be expensive. Fortunately, $C_\ell$ drops with $\ell$, $C_{\ell}\sim\ell^{-2}$, therefore very high $\ell$ modes 
			contribute very little and the summation can be truncated at $\ell\sim a\pi/\theta$, where $a\sim 5-10$ is a suitably chosen constant. We will see in section~\ref{ssec:iMaster_corr} 
			that one can also work with the binned quantities similar to the discussion for the pseudo-$C_\ell$ to further speed up the calculation. But first we discuss the estimator 
			used to measure the correlation functions.

		\subsection{Landy-Szalay estimator}
			\label{ssec:LS_estimator}
			The correlation functions measure the excess probability of finding galaxies around other galaxies at a given separation. These are measured by counting the excess number of pairs 
			of galaxies at a given separation relative to a distribution of randoms. The commonly employed 
			Landy-Szalay estimator \citep{Landy1993} is given by
			\begin{equation} 
				\xi(\theta_b)=\frac{B_\theta(DD-DR-RD+RR)}{B_\theta RR},
				\label{eq:LS_pairs}
			\end{equation}
			where $B_\theta$ is the binning operator, $D$ denotes the galaxies, $R$ denotes the randoms that correspond to the galaxy sample and \referee{ different two points, 
			e.g. $B_\theta DD$, denote the count of auto and cross pairs of galaxies and randoms with 
			separation $\theta$ that falls within the bin $\theta_b$.} It can be shown that when randoms follow the window function, LS 
			is an optimal estimator to compute the two point correlation functions \citep{Landy1993,Singh2017cov}.
			
			We now show that the LS estimator is equivalent to correlating the over density maps as defined in section~\ref{sec:maps}. The pairs counts can be written as integral over the 
			galaxy density field, i.e.
			\begin{equation} 
				DD(\theta)=\int_{\vtheta'} n_g(\vtheta')n_g(\vtheta'+\theta).
			\end{equation}
			$\int_{\vtheta'}=\int d\phi'd\theta'\sin(\theta')$ is the integral over the volume element in $\vtheta'$.
			From this we can show that the eq.~\eqref{eq:LS_pairs} is equivalent to 
			\begin{equation} 
				\xi(\theta_b)=\frac{B_\theta\int_{\vtheta'} (n_g(\vtheta')-n_R(\vtheta'))(n_g(\vtheta'+\theta)-n_R(\vtheta'+\theta))}{B_\theta\int_{\vtheta'} n_R(\vtheta'))n_R(\vtheta'+\theta)}.
			\end{equation}
			\refereetwo{Using the definition of $\delta$ from eq.~\eqref{eq:Pdelta_g} and ~\eqref{eq:Pdelta_g_R}, we get}
			\begin{equation} 
				\xi(\theta_b)=\frac{B_\theta\int_{\vtheta'} W(\vtheta')\delta_g(\vtheta')W(\vtheta'+\theta)\delta_g(\vtheta'+\theta)}{B_\theta \xi_W(\theta)},
				\label{eq:xi_W}
			\end{equation}
			where $\xi_W(\theta)$ is the correlation function of window.
			From eq.~\eqref{eq:xi_W}, the LS estimator is same as correlating the overdensity field, as long as the pixel size of the map is small enough to measure the minimum scale of interest. Computing 
			correlation functions via pixels can be cheaper when we have more than one galaxy (or more accurately, one random point) per pixel on average. Further gains can also be made by taking 
			advantage of the regularized nature of the pixel grid.
			Since use of randoms is simply a 
			monte-carlo method to account for window effects in the estimator, direct correlation of maps will also be free from additional noise introduced by the randoms.

			As discussed in section~\ref{sec:maps}, when computing
			correlation functions from galaxy catalogs, 
			usually one of following two approaches is adopted. In the first case, randoms are uniformly distributed on the sky and the galaxies are
			weighted by $1/W(\vtheta')$. This approach is commonly used \citep[e.g. ][]{Ross2012,Ross2020,Boss2016combined,Elvin-Poole2018} and 
			is equivalent to using the over density map as defined in eq.~\eqref{eq:delta_g}. In the second method, the random are weighted by the 
			window, i.e. $n_R(\vtheta')=\widebar{n}_g W(\vtheta')$ and galaxies are assigned uniform weight. This approach is equivalent to correlating the maps as defined in eq.~\eqref{eq:Pdelta_g}.
			In terms of optimality, same arguments as presented in section~\ref{sec:maps} applies to the question of weighing galaxies vs randoms, i.e. in general it is better to apply weights 
			to randoms.
			
			\refereetwo{It is also worth noting that the $B_\theta \xi_W(\theta)$ term (the $RR$ term) in the denominator is not strictly necessary if the window effects are properly accounted 
			for in the modeling. The use of this term in the estimator can also lead to the sub-optimal results as it up-weights the noisy modes where $B_\theta \xi_W(\theta)$ is low. It can be 
			more optimal to instead use a simple normailization constant, e.g. $A_W=f_\text{sky}$ (or $A_W=1$). In such a case, it can be easily shown using the equations in 
			appendix~\ref{app:cl_xi_scale_cuts} (replace $w(\theta)$ with $\xi_W$) that the correlation function and the pseudo-$C_\ell$ estimators are identical when the measurements are 
			performed over the full range of scales, $\theta\in[0,\pi]$ and $\ell\in[0,\infty]$. In the following section we relax this condition and study the relations between correlation 
			functions and power spectra over a limited range of scales.
			} 
				
		\subsection{{\sc $i$Master} for correlation functions}	
		\label{ssec:iMaster_corr}
		
		\refereetwo{ In the previous section, we have studied the correlation function estimators and showed their equivalence to the power spectra estimators. Now 
			we will study the relations between correlation 
			functions and power spectra for the practical cases where the measurements and the models are defined over a limited range of scales.
			We will generalize the {\sc $i$Master} algorithm to reconstruct power spectra from the correlation function measurements done over a limited $\theta$ range and 
			will also derive the expressions for properly implementing scale cuts on the correlation functions when the model is defined in Fourier space.
			}

			From eq.~\eqref{eq:xi_W} and appendix~\ref{app:xi}, the correlation function for a windowed field is given by
			\begin{equation}
				\xi(\theta_b)=\frac{B_{\theta}\xi_W(\theta)\mathcal H_{\theta,\ell}C_{\ell}} {B_{\theta}\xi_W(\theta)}.
				\label{eq:xi_W2}
			\end{equation}
			
			We can use the trick from section~\ref{ssec:iMaster_cl} and bin both $\xi$ and $C_\ell$ 
			\begin{equation} 
				\xi(\theta_b)=\frac{1}{B_\theta\xi_W(\theta)}B_{\theta}\xi_W(\theta)\mathcal H_{\theta,\ell}B^{-1}_{C} B_{C} C_{\ell},
			\end{equation}
			where $B_{\theta}$ is binning operator in $\theta$, acting on $\xi$ and $B_{C}$ is the binning operator in $\ell$, acting on $C_\ell$. Therefore we can write 
			the operation in terms of binned quantities, where the binned Hankel transform is given by
			\begin{equation}
				\mathcal H_{\theta_b,\ell_b}=\frac{1}{B_\theta\xi_W(\theta)}B_{\theta}\xi_W(\theta)\mathcal H_{\theta,\ell}B^{-1}_{C}.
			\end{equation}
			\refereetwo{As discussed in previous section~\ref{ssec:LS_estimator}, the ${B_\theta\xi_W(\theta)}$ term in the denominator of eq.~\eqref{eq:xi_W2} can be replaced with a constant 
				$A_W$, in which case the binned  Hankel transform changes to,
			\begin{equation}
				\mathcal H_{\theta_b,\ell_b}=\frac{1}{A_W}B_{\theta}\xi_W(\theta)\mathcal H_{\theta,\ell}B^{-1}_{C}.
			\end{equation}
			}
			
			$\theta_b$ is the effective $\theta$ at which the correlation function is  measured  and is given by
			\begin{equation}
				\theta_b=\frac{B_{\theta}\left(\theta\xi_W(\theta)\xi(\theta)\right)}{B_{\theta}\left(\xi_W(\theta)\xi(\theta)\right)}.
				\label{eq:thb}
			\end{equation}

			An aside, $\mathcal H_{\theta_b,\ell_b}$ can be used to transform $C_\ell$ to binned correlation function in $O(N_\text{bin}^2)$ time instead of $O(N_{\ell}N_\theta)$, where $N_\theta\gg 
			N_{bin}$ thereby speeding up the calculations without compromising the accuracy.
						
			Finally, we can invert Hankel transform to reconstruct the binned power spectra as 
			\begin{equation}
				C_{\ell_b}=\mathcal H_{\ell_b,\theta_b}^{-1}\xi(\theta_b).
			\end{equation}
			In general, in the presence of scale cuts in $\theta$, the inversion of $\mathcal H_{\ell_b,\theta_b}$ can be unstable. It is better to define inverse of 
			$\mathcal  H_{\ell_b,\theta_b}$ from the 
			definition of the inverse transform, where
			\begin{equation}
				C_{\ell}=2\pi\int d\theta\sin(\theta) _{s_1}d_{\ell,s_2}(\theta) \xi(\theta).
			\end{equation} 
			Replacing the integral with sum over discrete values of $\theta$, we can write
			\begin{equation}
				\mathcal H_{\ell,\theta}^{-1}=2\pi \Delta\theta\sin(\theta) _{s_1}d_{\ell,s_2}(\theta),
			\end{equation}
			which can then be binned to obtain
			\begin{equation}
				\mathcal H_{\ell_b,\theta_b}^{-1}=B_{C} \mathcal H_{\ell,\theta}^{-1} B_\theta^{-1}.
				\label{eq:hankel_binned}
			\end{equation}
			Notice that in this process we are reconstructing the true power spectra and not the pseudo-$C_\ell$ as long as the correlation functions are measured over the full range, $\theta\in[0,\pi]$. 
			The effects of the survey window are absorbed in the $\mathcal H_{\ell_b,\theta_b}^{-1}$ via the $B_\theta^{-1}$ operation.
			\refereetwo{ $B_\theta^{-1}$ is defined similar to eq.~\eqref{eq:B_inv} using the ratio $\xi(\theta)/\xi(\theta_b)$.}
			
			In practice, the correlation functions are usually measured over a limited range of scales, in which case we define
			\begin{equation}
				\xi_{cut}(\theta)=\xi(\theta)w(\theta), 
			\end{equation} 
			where $w(\theta)$ is a weight function (usually a top hat function with the limits of $\theta_{min},\theta_\text{max}$) that applies the scale cuts. Inverting $\xi_{cut}$,  we
			will get
			\begin{equation}
				\mathcal H_{\ell_b,\theta_b}^{-1}\xi_{cut}(\theta)=M_{\ell_b\ell_b'}C_{\ell_b'},
				\label{eq:Dl_xi}
			\end{equation} 
			where $M_{\ell_b\ell_b'}$ is the coupling matrix as defined in eq.~\eqref{eq:coupling_M} and eq~\eqref{eq:coupling_M_binned}, this time with $w(\theta)$ acting as window 
			( see appendix~\ref{app:corr_scale_cuts} for derivation).
			\refereetwo{Note that when $\theta$ range is small, the coupling matrix can be rather broad in which case separate binning operators on $B_D$ and $B_C$ may be required similar to 
			eq~\eqref{eq:coupling_M_binned}.}
			
			\begin{figure}
    	    		\centering
        	 		\includegraphics[width=\columnwidth]{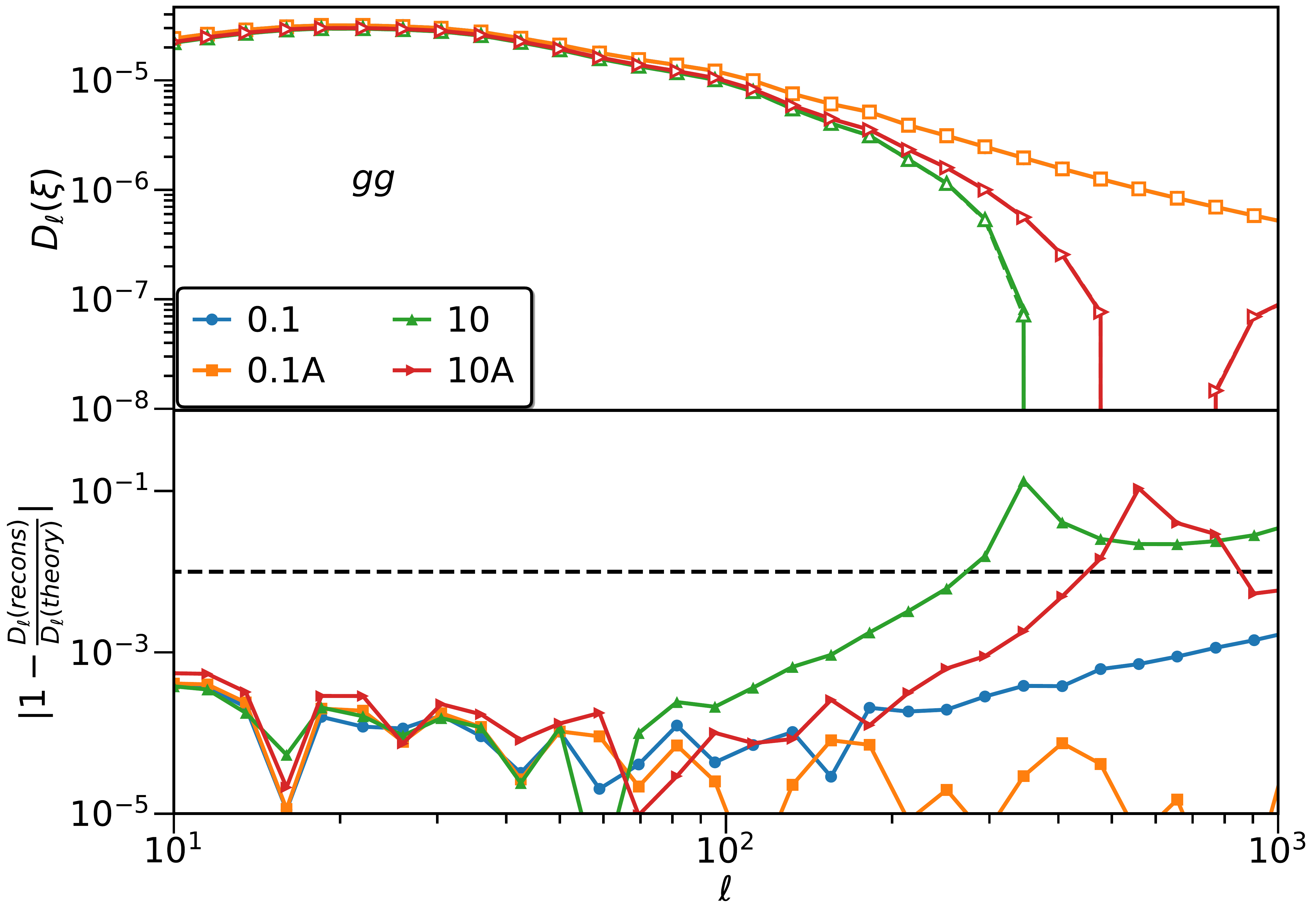} 
					\caption{Pseudo-$C_\ell$ power spectra obtained after inverse Hankel transform of the correlation function, $\xi$. $\xi$ was computed over the range $[0.01,600]$ 
					arcminutes. For blue and green points, I apply a hard cut on xi and 0.1 and 10 arcminutes respectively. Solid lines show the  $C_\ell'$ converted using coupling matrix, 
					$M_{\ell,\ell'}$. 
					Orange and Red point show the attempt to reduce the complexity of the coupling matrix by apodizing the $w(\theta)$ by
					multiplying its Hankel transform $w_\ell$ with a function that smoothly goes from one to zero in the range, $\ell\in[100,1000]$. While apodization helps,
					it does so by including information from lower $\theta$ than the original $\theta_\text{cut}$ and in practice one may need fairly strong apodization if $\xi$ is computed 
					over narrow $\theta$ range.
					}
					\label{fig:xi_Dl}
			\end{figure}
			In figure~\ref{fig:xi_Dl}, we see the $D_\ell$ obtained by the inverse Hankel transform of the correlation function. For a wide range, $\theta\in[0.1,600]$ arcminutes, the $D_\ell$ from inverse
			Hankel transform of correlation function and from convolving $C_\ell$ are consistent. However, for larger $\theta_{min}=10$ arcminutes, 
			the results from convolving $C_\ell$ are biased.
			This is because the coupling matrix $M_{\ell,\ell'}$ is very broad and the range $\ell\in[0,3000]$ used here for $C_\ell$ calculations is not enough.
			This is demonstration of the case where we require $\ell',\ell''\gg\ell$ when computing the coupling matrix (see discussion after eq.~\eqref{eq:coupling_M}).
			The figure also shows an attempt to reduce the complexity of the coupling matrix by apodizing the scale cut window $w(\theta)$. 
			Here a simple procedure was adopted where the $w_\ell$ is multiplied with a
			cosine function which goes from one to zero for $\ell\in [100,1000]$ and then $w_{\ell}$ is transformed back into $w(\theta)$. This apodized window effectively
			brings back some power from $\theta<10$ arcminutes and helps in partially reducing the bias. In practical applications, one will have to experiment with a few different 
			apodization schemes depending on the measurements being performed and scales being used to obtain good results.
									
			From eq.~\eqref{eq:Dl_xi}, the power spectra can be reconstructed from correlation functions as 
			\begin{equation}
				C_{\ell_b'}=M_{\ell_b'\ell_b}^{-1}\mathcal H_{\ell_b,\theta_b}^{-1}\xi_{cut}(\theta_b)=\mathcal M_{\ell_b',\theta_b}\xi_{cut}(\theta_b),
				\label{eq:iMaster_corr}
			\end{equation} 
			where we defined
			\begin{equation}
				\mathcal M_{\ell_b',\theta_b}=M_{\ell_b'\ell_b}^{-1}\mathcal H_{\ell_b,\theta_b}^{-1}.
			\end{equation}
			The covariance of reconstructed power spectra is 
			\begin{equation}
				\text{Cov}(C_{\ell_1},C_{\ell_2})=\mathcal M_{\ell_1,\theta_1} \text{Cov}(\xi_{\theta_1},\xi_{\theta_2})\mathcal M_{\ell_2,\theta_2}^T.
				\label{eq:imaster_cov}
			\end{equation} 
			
			\referee{Note that unlike existing methods in the literature, e.g. \cite{Joachimi2021}, the {\sc $i$Master} method we discuss here returns the unbiased $C_\ell$ power spectra at the 
			effective bin centers, $\ell_b$ and removes any effects of mode mixing from the correlation function measurements done over a limited range of scales. 
			The method also accounts for the window effects which has not been 
			done before in literature to the best of my knowledge. 
			As in section~\ref{ssec:iMaster_cl}, the results from eq.~\eqref{eq:iMaster_corr} and \eqref{eq:xi_cl_EE} can directly be compared with the model 
			$C_\ell$ predictions and do not require any additional operations \refereetwo{such as binning} or any corrections to be applied.
			}

		\begin{figure*}
			\begin{subfigure}[t]{\columnwidth}
    	    		\centering
        	 		\includegraphics[width=\columnwidth]{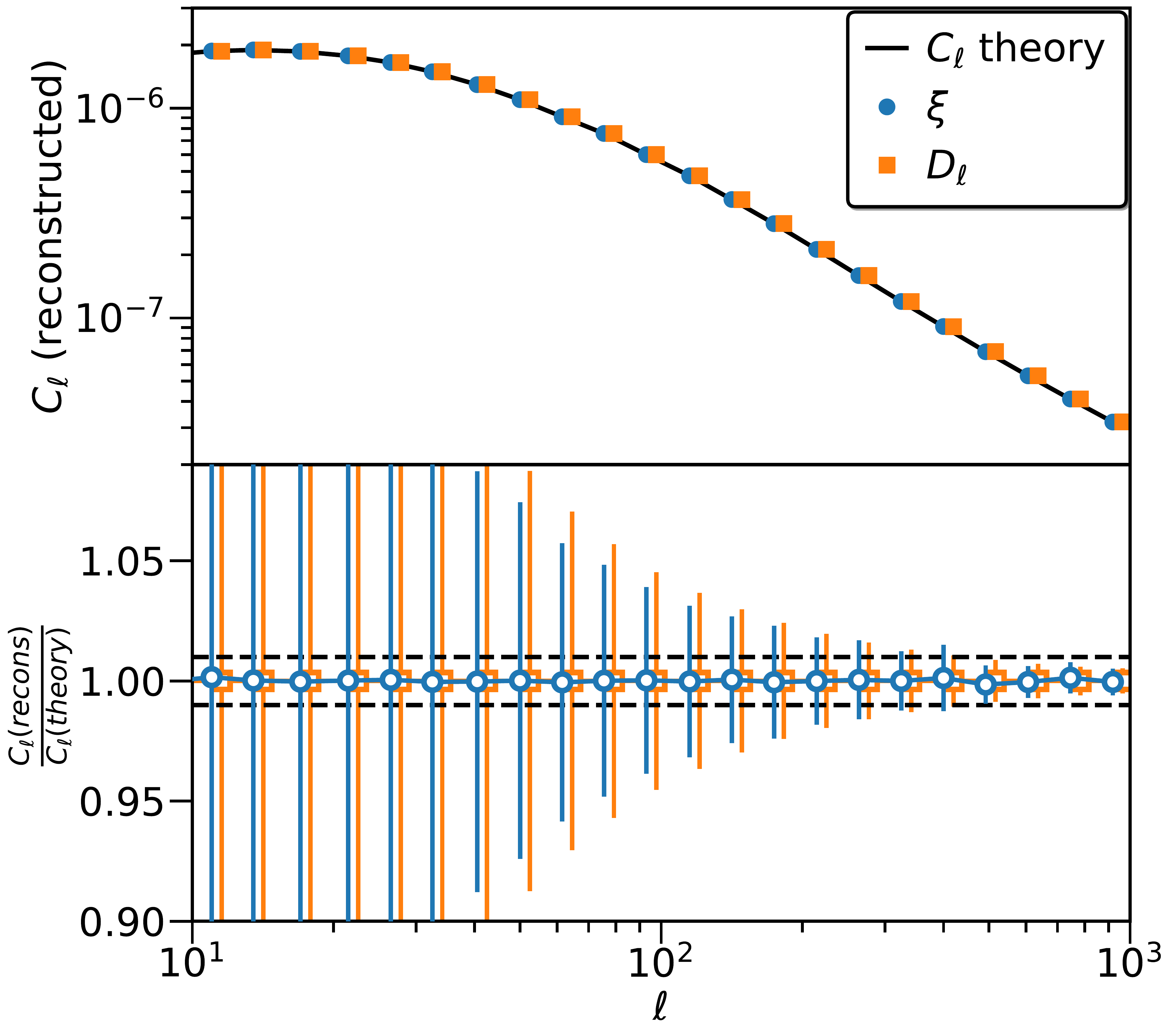}
					\caption{}
					\label{fig:}
			\end{subfigure}
			\begin{subfigure}[t]{\columnwidth}
    	    		\centering
        	 		\includegraphics[width=\columnwidth]{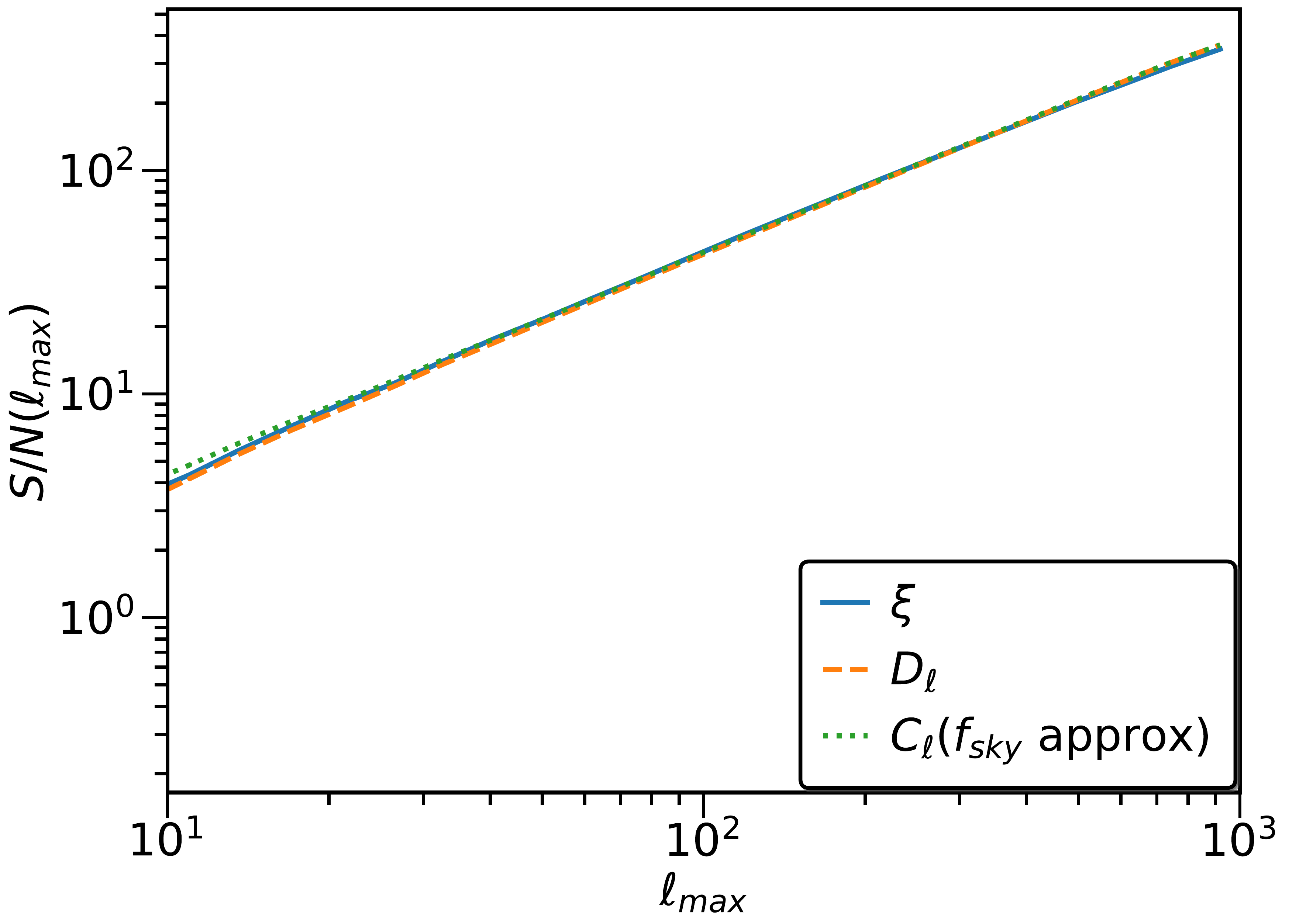}
					\caption{}
					\label{fig:}
			\end{subfigure}
			\caption{a) Power spectra reconstructed from binned correlation function (blue points) with $\theta\in[0.01,1200]$ arcminutes and the pseudo-$C_\ell$ power spectra from section
			\ref{ssec:iMaster_cl}. This plot uses $W_1$ window (see appendix~\ref{app:sims}). Lower panel shows the ratio with the input $C_\ell$, 
			with errorbars from the analytical covariance and one percent deviations marked by dashed black lines. Note that the $C_\ell$ reconstructed from  $\xi$ and $D_\ell$ are not 
			perfectly correlated due to the impact of limited $\theta$ range \refereetwo{and the window correction term in the correlation function estimator (see discussion in  
			section~\ref{ssec:LS_estimator} )}. 
			Different assumptions made in covariance calculations and the numerical noise in converting them also have some impact on the 
			errorbars shown and thus this comparison should be taken as approximate.
			b) Signal to noise ratio (S/N) as function of $\ell_\text{max}$ cut off for the reconstructed 
			power spectra in a). Also shown in the S/N for $C_\ell$ with a diagonal covariance with same $f_{sky}$. For $\ell_\text{max}>100$, all three curves agree to within $\sim\pm5\%$.}
			\label{fig:imaster_cl_xi_comp}
		\end{figure*}

		In fig.~\ref{fig:imaster_cl_xi_comp}, we see the comparison of the reconstructed power spectra from both correlation function and the direct pseudo-$C_\ell$ power spectra (all 
		calculations are analytical). Correlation functions are computed over a wide range, $\theta\in[0.01,1200]$ arcminutes, to lower the complexity of the coupling matrix. 
		Both correlation functions and pseudo-$C_\ell$ give 
		consistent results to well within $1\%$. The right panel of the figure also shows the comparison of the signal to noise ratio (S/N) as function of $\ell_\text{max}$. To obtain S/N of 
		reconstructed $C_\ell$, we inverted the analytical gaussian covariances of pseudo-$C_\ell$ \citep{Efstathiou2004} and correlation functions \citep[see appendix A of ][]{Singh2017cov} using the relation in 
		eq.~\eqref{eq:imaster_cov}. The S/N is defined as
		\begin{equation}
			S/N=\sqrt{C_\ell\text{Cov}^{-1}C_\ell}.
		\end{equation}
		Recently there has been some discussion about the apparent discrepancies or low correlation in the cosmological analysis from correlation functions and power spectra 
		\citep[e.g.][]{Doux2021,Hamana2020}. These discrepancies are known to be caused primarily by effects of the scale cuts imposed on the estimators, \refereetwo{in addition to some small 
		effects caused by approximations made in covariances and the estimators (see section~\ref{ssec:LS_estimator})}. 
		From eq.~\eqref{eq:xi_win_app} 
		and appendix~\ref{app:cl_xi_scale_cuts}, it can be shown
		that when full range of scales is used, the correlation function and power spectra have same information, i.e. up to the normalization factor of window in 
		correlation functions (the $RR$ term in the denominator), the correlation
		function and pseudo-$C_\ell$ are the same.
		Because of the impact of scale cuts, it is sometimes claimed that 
		correlation functions and power spectra provide complementary information. This is true in the technical sense as the coupling matrix becomes complex with scale cuts in $\theta$ (see 
		fig.~\ref{fig:xi_Dl}), leading to mixing of 
		information from larger $\ell$ range and the information from different $\ell$ modes is complementary 
		(assuming they are independent). However, such `complementarity' is not desirable if we do not have a good model for the subset of $\ell$ modes (which is usually the motivation for scale 
		cuts). Mixing of information from such scales
		only complicates the interpretation of the full posteriors from the analysis, even if the analysis is shown to be `unbiased' under some tests.
		To further understand the potential implications, consider the fact that the size of scatter shown in fig. 17 of \cite{Hamana2020} and fig. 9 of \cite{Doux2021} is comparable to the statistical 
		uncertainties on parameters and to the magnitude of tensions observed 
		between some of the weak lensing measurements and the predictions from Planck cosmology. 

		In general, if possible, one should keep data and model in the same space to avoid the complexities of transforming to the Fourier counterpart.
		Fig.~\ref{fig:imaster_cl_xi_comp} shows that when treated consistently, the correlation functions and power spectra carry similar information and the differences
		are primarily driven by scale cuts, which we need to implement carefully. We have addressed the implications of scale cuts imposed on correlation functions when reconstructing 
		the power spectra and we will address the inverse of this process, namely the impact on correlation function when the scale cuts are imposed on the model in the Fourier space in 
		section~\ref{sssec:xi_convolve}. Before that, we now address the issue of reconstructing lensing E/B modes from the cosmic shear correlation functions.
		
		\subsubsection{E/B mode reconstruction}
		\label{ssec:EB_recons}
		\refereetwo{
			For cosmic shear measurements, we typically measure two sets of correlation functions, $\xi_+$ and $\xi_-$, which can be written in terms of the underlying EE and BB power spectra 
			as
			\begin{align}
				\xi_\pm(\theta_b)=H_{\pm,\theta_b,\ell_b}(C_\ell^{EE}\pm C_\ell^{BB}),
			\end{align}
			where $H_{\pm,\theta_b,\ell_b}$ are Hankel transform operators as defined in eq~\eqref{eq:hankel_binned} with spin-2 wigner-d matrices, $_{2}d_{\ell,\pm2}$.}
			
			\refereetwo{
			Using eq.~\eqref{eq:iMaster_corr} we can also do a clean E/B mode separation for the case of cosmic shear, by converting  $\xi_{\pm}$ to pseudo-$C_\ell$ as
			\begin{align}
				D_{\ell_b}^\pm=\mathcal H_{\pm,\ell_b,\theta_b}^{-1}\xi_{\pm}(\theta_b).
			\end{align} 
			Here we assume that the $\xi_\pm$ are measured over the same $\theta$ range.
			Now we can obtain E/B pseudo-power spectra as
			\begin{align}
				D_{\ell_b}^{EE}=\frac{1}{2}(D_{\ell_b}^+ + D_{\ell_b}^-)\label{eq:xi_Dl_EE}\\
				D_{\ell_b}^{BB}=\frac{1}{2}(D_{\ell_b}^+ - D_{\ell_b}^-)\label{eq:xi_Dl_BB}.
			\end{align} 
			$D_{\ell_b}^{EE,BB}$ and the coupling matrices are defined in section~\ref{ssec:pcl}.
			From $D_{\ell_b}^{EE,BB}$, we can reconstruct the underlying E/B power spectra as 
			\begin{align}
				\widehat{C}_{\ell_b}^{EE}&=\left(M^{+EE}_b\right)^{-1}\widehat{D}_{\ell_b'}^{EE}-\left(M^{+EE}_b\right)^{-1}M^{-BB}_bC_{\ell_b}^{BB}\label{eq:xi_cl_EE}\\
				\widehat{C}_{\ell_b}^{BB}&=\left(M^{+BB}_b\right)^{-1}\widehat{D}_{\ell_b'}^{BB}-\left(M^{+BB}_b\right)^{-1}M^{-EE}_bC_{\ell_b}^{EE}\label{eq:xi_cl_BB}.
			\end{align}
			$M_b$ are binned coupling matrices, with the superscript, e.g. $+EE$, referring to the power spectra used in defining the $B^{-1}$ operator. 
			In eq.~\eqref{eq:xi_cl_EE} and \eqref{eq:xi_cl_BB}, we 
			have also made a distinction between the quantities measured from data, denoted with $\widehat{}$, and the ones estimated from theory. We are using the model $C_{\ell_b}^{EE,BB}$
			to subtract out the leakage contribution of the form, $M^-C_{\ell_b}$. This is likely to be sufficient for most applications as $(M^+)^{-1}M_-$ is typically small 
			($\lesssim10^{-2}$) for broad windows.
			When noise is sub-dominant, it may be desirable to replace the $C_\ell$ from model with the ones from data in order to cancel sample variance, 
			e.g. in the case where we are attempting to detect and reconstruct small $B$ mode power spectra.
			Under such a scenario, we can use an iterative method where the initial estimates of $\widehat{C}_{\ell_b}^{EE,BB}$ from eq.~\eqref{eq:xi_cl_EE} and \eqref{eq:xi_cl_BB} 
			can be used to replace theory $C_{\ell_b}^{EE,BB}$ in the next iteration. Only few such iteration will be required as the excess sample variance effects from leakage 
			($\widehat{C}_\ell-C_\ell$) 
			scale down as $[(M^+)^{-1}M_-]$. We have also neglected the possibility of reconstructing power spectra from terms involving $M^{-}$ terms, i.e., reconstructing $
			\widehat{C}_{\ell_b}^{EE,BB}$ from $\widehat{D}_{\ell_b}^{BB,EE}$ (EE from BB and BB from EE). In such reconstruction the noise scales as $[(M^-)^{-1}M_+]$, which is very large 
			($\gtrsim10^{2}$) when window is broad (narrow in $\ell$ space) and thus we do not expect to gain much in terms of signal to noise of the overall $\widehat{C}_{\ell_b}^{EE,BB}$.
			}
			
			\refereetwo{
			Fig.~\ref{fig:corr_EB} demonstrates the reconstruction of the EE and BB power spectra reconstructed from $\xi_\pm$ correlation function. Here we assumed $C_\ell^{BB}=0.1C_\ell^{EE}$ 
			to demonstrate the steps in the reconstruction (we assume $C_\ell^{BB}=0$ in rest of the paper). The pseudo-$C_\ell$ obtained by inverse Hankel transform of correlation functions are 
			biased by few percent as expected. Reconstructing the power spectra without the correction for leakage, i.e. we set $M^-C_\ell^{EE,BB}=0$, also leads to biased results, especially for 
			the B mode power spectra which is smaller and hence more sensitive to these biases. Using expressions in eq.~\eqref{eq:xi_cl_EE} and \eqref{eq:xi_cl_BB} allow us to obtain unbiased 
			results to better $1\%$ accuracy (note that this accuracy depends on the choice of scale cuts on $\xi_\pm$ and the $\ell_\text{max}$ of coupling matrix. Smaller $\theta$ range in 
			correlation functions will require larger $\ell_\text{max}$ for same accuracy). Setting $M^-C_\ell^{EE,BB}=0$ initially and then adopting the iterative procedure as discussed above 
			converges to same unbiased results in two iterations after which there are no further improvement (not shown).
			\begin{figure}
    	    		\centering
        	 		\includegraphics[width=\columnwidth]{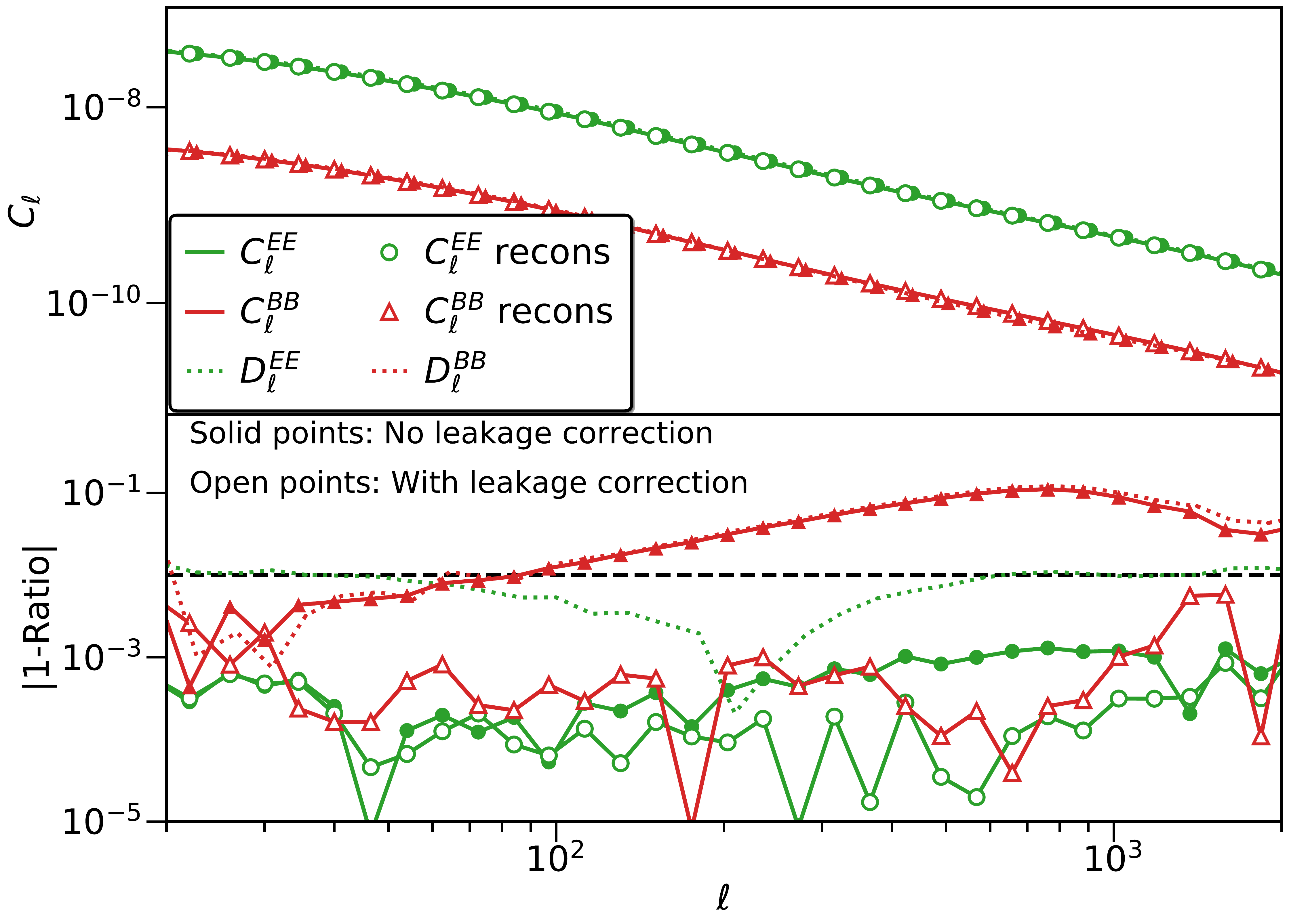}
					\caption{\refereetwo{Demonstration of shear EE and BB power spectra reconstruction from $\xi_\pm$ correlation functions via the {\sc $i$Master} algorithm. \emph{Upper panel}: Solid lines 
						show the true underlying power spectra, where we assumed $C_\ell^{BB}=0.1C_\ell^{EE}$ for this demonstration. Dotted lines show the pseudo-power spectra constructed via
						inverse Hankel transform, solid points show reconstructed power spectra from $D_\ell$ while setting leakage terms $M^-C_\ell^{EE,BB}=0$ 
						(see eq.~\eqref{eq:xi_cl_EE} and \eqref{eq:xi_cl_BB}) and open points show the results when correct  $M^-C_\ell^{EE,BB}$ is used. \emph{Lower panel}: Fractional errors in 
						different curves with respect to the true $C_{\ell}^{EE,BB}$. Both leakage and the window effects lead to errors of order few percent which are removed when using correct
						expressions in eq.~\eqref{eq:xi_cl_EE} and \eqref{eq:xi_cl_BB}. The residuals errors are primarily from the limited $\theta$ and $\ell$ range used in the tests and can be 
						reduced further by expanding these ranges. }
					}
					\label{fig:corr_EB}
			\end{figure}
			}
						
		\subsubsection{Model cuts in Fourier space}
		\label{sssec:xi_convolve}
			Many of the cosmological models are written and validated in the Fourier space and thus have a well defined cuts in the Fourier space based on their scales of validity. 
			This implies truncation in the summation in 
			eq.~\eqref{eq:xi0}. Similar to the issues we addressed in the case of power spectra, the truncation in the Fourier counterpart  
			should result in a convolution on the correlation functions of the form \referee{(see appendix~\ref{app:corr_scale_cuts} for derivation)},
			\begin{equation} 
				\xi_{\ell-\text{cut}}(\theta)=\mathcal H_{\theta,\ell} C_{\ell}b_{\ell}=b(\theta,\theta')\circledast\xi(\theta'),
				\label{eq:xi_cut_convolve}
			\end{equation}
			where $b_{\ell}$ is the truncation function in the Fourier space and
			in the second equation we wrote the computed correlation function
			as a convolution between true underlying correlation function and the Hankel transform of $b_\ell$, given by (see appendix~\ref{app:corr_scale_cuts})
			\begin{align}
				b(\theta,\theta')=\sum_{\ell} b_\ell \frac{2\ell+1}{4\pi} {_{s_1}Y_{\ell,{s_2}}}(\theta)_{s_1}Y_{\ell,{s_2}}(\theta').
			\end{align}
			Therefore in the case of a theory cut off defined in Fourier space, we can simply convolve the measured correlation function with the $b(\theta,\theta')$ to impose the `scale cuts' 
			before running the inference chain.
			Using our binning trick from previous sections, we can define the binned version of the coupling matrix,
			\begin{align}
				b(\theta_b,\theta_b')=B_\theta b(\theta,\theta') B_\theta^{-1}.
			\end{align}
			Also note that $b_\ell$ is operationally same as an isotropic pixel/beam smoothing applied on over density maps with the beam function given by $\sqrt{b_\ell}$. Thus this 
			method can be used to account for such smoothing effects as well.
			We can also account for correct factors of $b_\ell$ in the covariance and cross covariance calculations by considering each tracer to be smoothened by its own `beam' given by 
			$\sqrt{b_\ell}$, which can be different for different tracers (also remember that shot noise in covariance will no longer scale as $1/N_\text{pairs}$ after smoothing).

			While $b_\ell$ is typically chosen to be top hat function, it can lead to convolution over rather large scales in $\theta$. A choice of with more compact 
			$b(\theta,\theta')$ can be given by a function that drops more smoothly from one to zero, such as 
			\begin{align}
				b_{\ell,\cos}=\begin {cases} 
				1,  &\ell<= \ell_\text{cut,min}\\
				\cos{(\frac{\pi}{2} \frac{\ell-\ell_\text{cut,min}}{\ell_\text{cut,max}-\ell_\text{cut,min}}) },  &\ell_\text{cut,min}<\ell<\ell_\text{cut,max}\\
				0  &\ell>= \ell_\text{cut,max}
				\end{cases},
			\end{align}
			where $b_{\ell,\cos}$ use a cosine function to smoothly truncate the power spectra to zero between $\ell_\text{cut,max}-\ell_\text{cut,min}$. A larger separation between 
			$\ell_\text{cut,max}$ and $\ell_\text{cut,min}$ will lead to narrower convolution in $\theta$ space.

			More frequently, the cutoff in theoretical models are defined in the comoving ($k$) space, in which case the $b_\ell$ can be written as
			\begin{equation}
				b_{\ell,\text{k-cut}}=\frac{C_{\ell,\text{k-cut}}}{C_\ell},
			\end{equation}
			where $C_{\ell,\text{k-cut}}$ is computed with cut off in $k$ and $C_\ell$ is true underlying power spectra without any cutoff. This requires us to have some estimate of 
			$C_\ell$ and in practice it will likely be better to use some combination of $b_{\ell,\text{k-cut}}$ and $b_{\ell,\cos}$, i.e. $b_{\ell,\cos}\times b_{\ell,\text{k-cut}}$, for more
			accurate results.
			
			\begin{figure}
    	    	\centering
	        	 \includegraphics[width=\columnwidth]{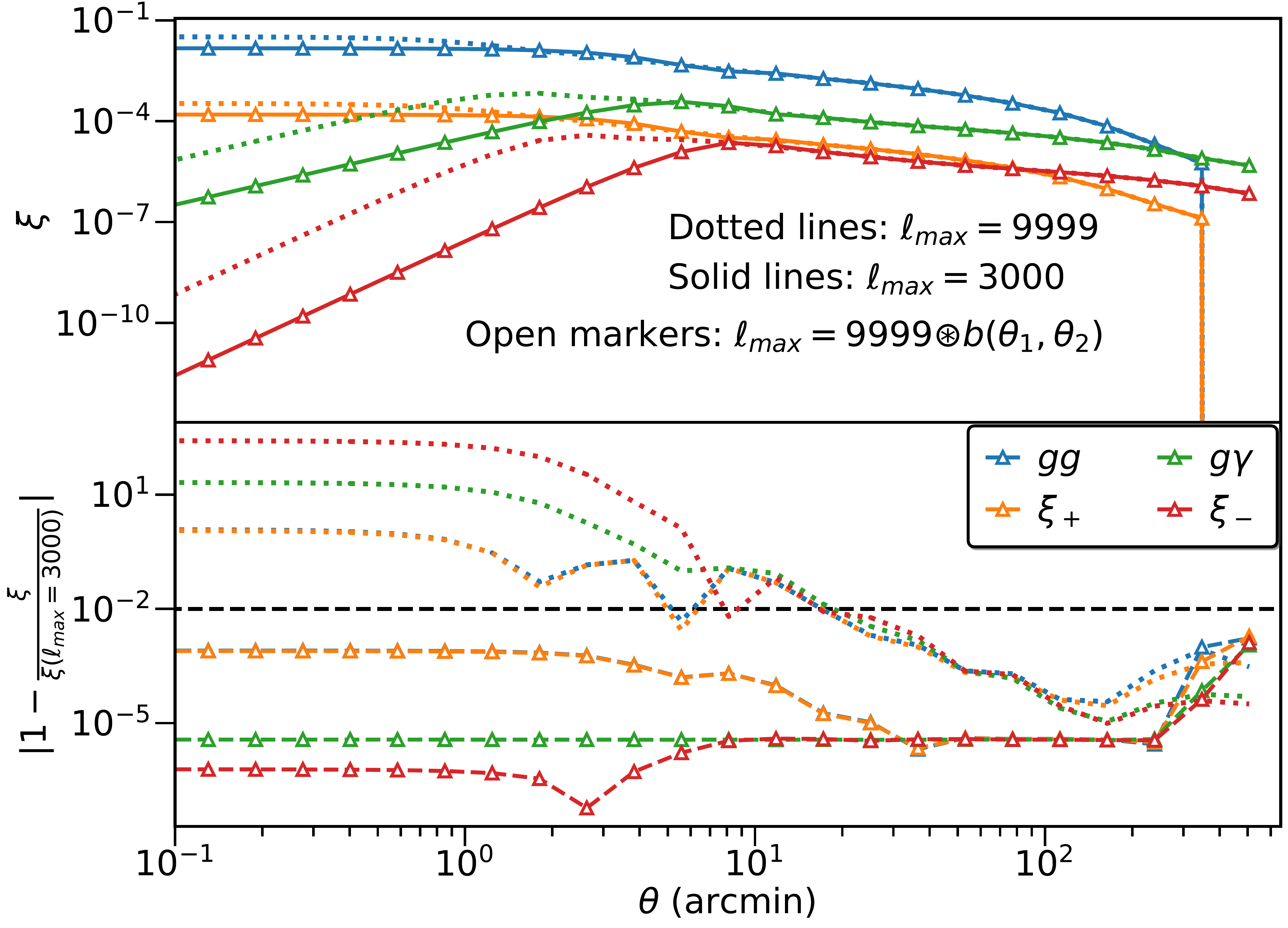}
				\caption{Comparison of correlation functions obtained with different $\ell$ cuts for auto and cross correlations of galaxies and shear. 
						Dotted lines show $\xi$ obtained with $\ell_\text{max}=10^4$, while solid lines show the case with a theory cut applied using $b_\ell$, with $\ell_\text{max}=3000$. 
						\refereetwo{The open points show the results obtained by convolving $\xi$ obtained using $\ell_\text{max}=10^4$ with the $b(\theta,\theta')$ to reproduce 
						the $\xi$ with $\ell_\text{max}=3000$ cut. The lower panel shows the fractional differences. Dashed lines are for the ratio of $\xi$s with $\ell_\text{max}=10^4$ and $\ell_\text{max}
						=3000$ while dashed lines with open markers show the same, using $\xi$ with $\ell_\text{max}=10^4$ and convolved with $b(\theta,\theta')$.
						}
				}
				\label{fig:xi_convolve}
			\end{figure}
			
		Fig.~\ref{fig:xi_convolve} shows the effects of convolving the correlation function with $b(\theta,\theta')$ to account for the effects of the scale scale cuts imposed on the 
		model. After convolution, the correlation function with high $\ell_\text{max}$ agrees to much better than $1\%$ with the correlation function with lower $\ell_\text{max}$ cut.
		
		\referee{For the case of galaxy-shear cross correlations, it is also worth comparing the effects of eq.~\eqref{eq:xi_cut_convolve} to the $\Upsilon$ 
		estimator suggested by \cite{Baldauf2010}. The $\Upsilon$ estimator is relevant for applying scale cuts when the model and hence the scale cuts are described in the real
		space. Eq.~\eqref{eq:xi_cut_convolve} should be used when the model and its cuts are described in the Fourier space.
		}
			
		We can also write convolution operators for the case of flat-sky approximation, for both projected and three dimensional (spectroscopic) statistics \citep{Baddour2014}
			\begin{align}
				b(\theta,\theta')_\text{flat sky}&=\int d\ell \ell b_\ell J_n(\ell\theta)J_n(\ell\theta'),\\
				b(r,r')&=\int dk k^2 b(k) j_n(kr)j_n(kr').
			\end{align}

			
	\subsubsection{Hybrid cuts}
		\refereetwo{In practice, both model and data may have scale cuts applied which may complicate the analysis when data and model are in separate spaces. For example, we may have
		correlation function measurements, with some scale cuts due to systematics such as blending or fiber collisions while the model is defined in the fourier counterpart, i.e. power 
		spectra. In such a case, eq.~\eqref{eq:xi_cut_convolve} modifies to
		\begin{equation} 
				\xi_{\ell,\theta-\text{cut}}(\theta)=b(\theta,\theta')\circledast\xi(\theta')w(\theta')=\mathcal H_{\theta,\ell} D_{\ell}b_{\ell},
				\label{eq:xi_ltcut_convolve}
			\end{equation}
		where in the second part of the equation we now have pseudo-$C_\ell$ instead of $C_\ell$.}
		
		\refereetwo{
		In such a case, it is easiest to work with the reconstructed power spectra via {\sc $i$Master}, as in eq.~\eqref{eq:iMaster_corr} or in eq.~\eqref{eq:Master_inv} (with $M$ from 
		eq.~\eqref{eq:coupling_M_binned}) if one is working with the pseudo-$C_\ell$ measurements. For E/B power spectra, the procedure in section~\ref{ssec:EB_recons} should be followed.
		}

		\refereetwo{ It can also be computationally efficient to combine the correlation function and pseudo-$C_\ell$ estimators for the reconstruction of power spectra on all scales.
		The pair counting correlation functions estimators are fast to implement at small scales (total number of pairs is small) while the pseudo-$C_\ell$ estimators are faster at large scales (Fourier transforms can be computed on coarser grids/maps). Reconstructing $C_\ell$ from
		both estimators and then combining them with minimum variance weighting can result in faster reconstruction of $C_\ell$ over a broad range of modes.
		}
					
	\subsection{Window errors}
			As discussed in section~\ref{ssec:LS_estimator}, the correlation function measured using the LS estimator is given by 
			\begin{equation} 
				\xi(\theta_b)\approx\frac{B_\theta \xi_W(\theta)\xi(\theta)}{B_\theta \xi_W(\theta)},
				\label{eq:xiW3}
			\end{equation}
			where $\xi_W(\theta)$ is the correlation function of the window. Note that the binning operation is applied separately to the numerator and the denominator. 
			If $\xi_W(\theta)$ has a strong gradient within the bin, it can introduce a bias, namely that the effective scale of the measurement $\theta_b$ (eq.~\eqref{eq:thb}) 
			will be different from the bin center computed without accounting for the window effect.

			Since the Landy-Szalay estimator uses the estimated window (randoms follow the estimated window function), which maybe different from the true underlying window, 
			correlation functions also suffer from the window bias. 
			The numerator of eq.~\eqref{eq:xiW3} has the correlation function of the true window ($W$) while the 
			denominator has the correlation function of the estimated window($\widehat{W}$). In
			presence of these biases we get
			\begin{equation} 
				\xi(\theta_b)=\frac{B_\theta \xi_W(\theta)\xi(\theta)}{B_\theta (1+m_\theta(\theta))\xi_W(\theta)}+a(\theta_b),
			\end{equation}
			where $a(\theta_b)$ is the correlation function of the additive bias and we introduced $m_\theta$ to refer to the multiplicative bias in the correlation function of the window. 
			As in the power spectra, $a(\theta_b)$ is easily accounted for by using the correlations method \citep[e.g.][]{Ross2012}, but $m(\theta)$ is usually harder to estimate 
			and can introduce significant biases, especially if it has a strong scale dependence. Since correlation functions and pseudo-$C_\ell$ are same, we can use the same methods as described in 
			section~\ref{ssec:Dl_error} to understand these window biases and add corrections terms to the data/model or to the covariance.

\section{Conclusion}
	\refereetwo{
			\begin{algorithm}
			\caption{{\sc $i$Master} for power spectra}
			\label{alg:imaster}
			\begin{algorithmic}
				\item {\bf Input}: Measured power spectra $\widehat{D}_\ell$; Window power spectra $\widehat{W}_\ell$; Model power spectra $C_\ell$, $N_\ell$; Binning operators $B_D$, $B_C$.
				\STATE 1. Subtract the noise from $\widehat{D}_\ell$, i.e. $\widehat{D}_\ell=\widehat{D}_\ell-N_\ell$.
				\STATE 2. Bin $\widehat{D}_\ell$, to obtain $\widehat{D}_{\ell_b}=B_D \widehat{D}_\ell$.
				\STATE 3. Bin ${C}_\ell$, to obtain ${C}_{\ell_b'}=B_C \widehat{C}_{\ell'}$.
				\STATE 4. Obtain $\ell_b'$, eq.~\eqref{eq:ell_b}.
				\STATE 5. Obtain $B_C^{-1}$, eq.~\eqref{eq:B_inv}.
				\STATE 6. Compute the coupling matrices, $M_{\ell,\ell'}$, eq.~\eqref{eq:coupling_M}.
				\STATE 7. Bin the coupling matrices to obtain $M_{\ell_b,\ell_b'}$, eq.~\eqref{eq:coupling_M_binned}.
				\STATE 8. Compute the pseudo-inverse of $M_{\ell_b,\ell_b'}$ and obtain $\widehat{C}_{\ell_b'}= M_{\ell_b',\ell_b}^{-1}\widehat{D}_{\ell_b}$.
				\STATE For E/B mode separation, use eq.~\eqref{eq:xi_cl_EE} and \eqref{eq:xi_cl_BB} as discussed in section~\ref{ssec:EB_recons}. Use iterative method if needed.
				\item {\bf Output}: Binned power spectra $\widehat{C}_{\ell_b'}$ and the effective $\ell$ values for the bin $\ell_b'$.
			\end{algorithmic}
			\end{algorithm}}
			\refereetwo{
			\begin{algorithm}
			\caption{{\sc $i$Master} for correlation functions}
			\label{alg:imaster_corr}
			\begin{algorithmic}
			\item {\bf Input}: Measured correlation functions $\widehat{\xi}(\theta_b)$; Window correlation functions, $\widehat{\xi}_W(\theta)$ and $\widehat{\xi}_W(\theta_b)$; 
							   Model power spectra $C_\ell$; Binning operators $B_\theta$, $B_C$ ($B_D$ if required); $w(\theta)$ defining the scale cuts.
			\item {\bf Output}: Binned power spectra $\widehat{C}_{\ell_b}$.
			\STATE 1. Compute the Hankel transform operator and its inverse.
			\STATE 2. Compute the binned theory $\xi(\theta_b)$ using eq.~\eqref{eq:xi_W2} and obtain $B_\theta^{-1}$.
			\STATE 3. Bin the inverse Hankel transform operator, eq.~\eqref{eq:hankel_binned}.
			\STATE 4. Obtain pseudo-$C_\ell$ power spectra, eq.~\eqref{eq:Dl_xi}.
			\STATE    For cosmic shear E/B power spectra, use eq.~\eqref{eq:xi_Dl_EE} and \eqref{eq:xi_Dl_BB}.
			\STATE 5. Obtain $W_\ell$ using inverse Hankel transform of $w(\theta)$.
			\STATE 6. Go to step 3. of the algorithm 1: {\sc $i$Master} for power spectra.
			\end{algorithmic}
			\end{algorithm}
			}

	In this paper we reviewed the formalism for the measurement and modeling of two point functions of the cosmological tracers, 
	starting from the process of making maps from the catalogs to the measurements
	of the power spectra and the correlation functions and addressed several of the issues related to the importance of survey window, biases in its modeling as well as the issues relevant to 
	imposing scale cuts.
	
	In section~\ref{sec:maps}, we derived the expressions for the over density field and the expressions for shot noise contributions to the two different estimators. We showed that the window 
	effectively acts as the inverse noise weight and hence window weighted estimator has in general lower noise compared to the estimator in which the window is removed. In the noise dominated regime, 
	this estimator is nearly equivalent to the quadratic estimator (FKP) for the gaussian field and thus for many LSS surveys, which are still noise dominated on most scales, is close to 
	optimal. 
	While the focus of discussion was using maps, in sections~\ref{sec:maps} and \ref{ssec:LS_estimator}, we also discussed that applying systematics weights on galaxies, as is commonly done, 
	is equivalent of the suboptimal estimator and it is better to apply such weights on the randoms. It is also worth remembering that the arguments we presented are equally valid along 
	the line of sight (redshift) direction and in fact most of the studies in the literature apply the window weighting along the redshift direction.
	The $\frac{dn}{dz}$ weights that enter the $C_\ell$ model calculations are the redshift window weights and in the spectroscopic analysis BOSS collaboration \citep[e.g. ][]{Ross2020,Boss2016combined} 
	applied the FKP weights along the redshift space. 
	
	We also discussed the impact of the window estimation on the various estimators. In terms of the over density field, both estimators we discussed in section~\ref{sec:maps} are 
	affected similarly by the biases in window estimation and using the sub-optimal estimator does not in any way help with the problem of window modeling (even in the absence of biases, one 
	still need to model the mask which is not much easier). In section~\ref{ssec:Dl_error} we studied the impact of these window biases on the estimation of the pseudo-$C_\ell$ power spectra.
	Window biases lead to both additive and multiplicative biases in the power spectra. Additive biases have received some attention in the literature 
	\citep[e.g.][]{Ross2012,Leistedt2013}. Multiplicative biases while higher order, can be equally concerning for measurements with O(1\%) precision requirements. We also saw that these 
	biases can be more important at smaller scales (high $\ell$) which contribute a large fraction of the cosmological information.
	
	It is also worth stressing that these issues are equally applicable to tracers such as galaxy shear. In the case of shear, we normally think of multiplicative and additive biases on the
	ensemble bases. However, if these biases vary with the photometric conditions of the survey, then they can also be thought of as the part of the window and can be modeled using the 
	methods in this paper and elsewhere in the literature.
	
	In section~\ref{sec:power_spectra} we studied the estimator for the measurement of the power spectra, namely the pseudo-$C_\ell$ estimator and the algorithms used in the modeling. 
	We discussed the pesky issues involved in estimating the coupling matrices which can get quite complex for a non trivial window or for a power spectra that falls slowly 
	or is nearly flat. For such cases the window needs to be estimated accurately out to very high $\ell$, consistent with our discussion on systematics. In terms of algorithms, we saw that the 
	standard {\sc Master} algorithm can be biased due to incorrect assumptions about the scaling of the power spectra and our improved {\sc $i$Master} algorithm corrects for those
	biases by using the correct power spectra from the model. \referee{This algorithm is more powerful than the existing algorithms \citep[e.g. ][]{Alonso2019} as it allows direct comparison with
	\refereetwo{unbinned} theoretical models computed at effective bin centers, $\ell_b$ (i.e. binning is performed analytically) and does not require any corrections to be applied.
	}\refereetwo{Furthermore, the algorithm is also more optimal as it allows for cleaner extraction of information for a given set of scales by undoing the effects of mode mixing in the pseudo-$C_\ell$ estimator. We list the steps involved in {\sc $i$Master} computation in the algorithm~\ref{alg:imaster}.}

	In section~\ref{sec:correlation_function}, we generalized the {\sc$i$Master} algorithm to the correlation functions. Using this algorithm it is possible to reconstruct the 
	power spectra from the correlation functions, including the E/B mode separation, though it can be hard (coupling matrix is complex) if the range of scales at which correlation function is 
	measured is limited.
	\referee{Similar to the case of pseudo-$C_\ell$ mentioned earlier, power spectra modes reconstructed from correlation functions via {\sc$i$Master} can be directly compared with
	\refereetwo{unbinned} theoretical models computed at effective bin centers (i.e. binning+window corrections are performed within the algorithm) and do not require any corrections to be 
	applied.
	} \refereetwo{ Steps involved in reconstruction of power spectra from correlation functions are listed in the algorithm~\ref{alg:imaster_corr}.}
	We also developed the proper method to convolve the correlation functions in order to account for the scale cuts on the model in the Fourier space. This method prevents the leakage of 
	information from
	the scales in Fourier space that are not properly modeled.
	
	The {\sc$i$Master} algorithm is also useful in speeding up the computations (during sampling) as after the initial setup, the computational complexity is reduced to $O(N_\text{bin}^2)$ 
	instead of 
	$O(\ell_\text{max}^2)$. That being said, the performance of these algorithms becomes even more important in the memory management during a large analysis, 
	such as an LSST like $3\times2$ analysis, where the 
	amount of peak memory requirement for the analysis decreases from $O(N_\text{corr}\times\ell_\text{max}^2)$ to $O(N_\text{corr}\times N_\text{bin}^2)$, where $N_\text{corr}\sim O(100)$ is 
	the number of 
	correlation pairs. For these large analysis, it will become imperative to use such an algorithm for computationally fast and efficient sampling of the large parameter spaces.

\section*{Availability of data \& code}
No new data were generated or analyzed in support of this research.

The code used for the computations in this paper is available at \url{https://github.com/sukhdeep2/Skylens_public/tree/imaster_paper/}.

\section*{Acknowledgements}
I would like to thank Yin Li for many stimulating discussions that have helped me in this work and for providing detailed feedback on this manuscript. I also thank James Sullivan, Tanveer Karim and Mehdi 
Rezaie for helpful discussions and feedback on this work.
I also thank the anonymous referee for several helpful comments and questions that helped in improving the presentation and explanations in the paper.

I am supported by the McWilliams fellowship at the Carnegie Mellon university. 
Part of this work was done at University of California, Berkeley, where I was supported by a postdoctoral fellowship at the Berkeley center for cosmological physics.

\bibliographystyle{mnras}
  \bibliography{sukhdeep_im}

\begin{thebibliography}{}
\makeatletter
\relax
\def\mn@urlcharsother{\let\do\@makeother \do\$\do\&\do\#\do\^\do\_\do\%\do\~}
\def\mn@doi{\begingroup\mn@urlcharsother \@ifnextchar [ {\mn@doi@}
  {\mn@doi@[]}}
\def\mn@doi@[#1]#2{\def\@tempa{#1}\ifx\@tempa\@empty \href
  {http://dx.doi.org/#2} {doi:#2}\else \href {http://dx.doi.org/#2} {#1}\fi
  \endgroup}
\def\mn@eprint#1#2{\mn@eprint@#1:#2::\@nil}
\def\mn@eprint@arXiv#1{\href {http://arxiv.org/abs/#1} {{\tt arXiv:#1}}}
\def\mn@eprint@dblp#1{\href {http://dblp.uni-trier.de/rec/bibtex/#1.xml}
  {dblp:#1}}
\def\mn@eprint@#1:#2:#3:#4\@nil{\def\@tempa {#1}\def\@tempb {#2}\def\@tempc
  {#3}\ifx \@tempc \@empty \let \@tempc \@tempb \let \@tempb \@tempa \fi \ifx
  \@tempb \@empty \def\@tempb {arXiv}\fi \@ifundefined
  {mn@eprint@\@tempb}{\@tempb:\@tempc}{\expandafter \expandafter \csname
  mn@eprint@\@tempb\endcsname \expandafter{\@tempc}}}

\bibitem[\protect\citeauthoryear{{Abazajian} et~al.,}{{Abazajian}
  et~al.}{2016}]{CMB_S4}
{Abazajian} K.~N.,  et~al., 2016, arXiv e-prints, \href
  {https://ui.adsabs.harvard.edu/abs/2016arXiv161002743A} {p. arXiv:1610.02743}

\bibitem[\protect\citeauthoryear{{Abbott} et~al.,}{{Abbott}
  et~al.}{2020}]{DESY1_cluster}
{Abbott} T.~M.~C.,  et~al., 2020, \mn@doi [\prd] {10.1103/PhysRevD.102.023509},
  \href {https://ui.adsabs.harvard.edu/abs/2020PhRvD.102b3509A} {102, 023509}

\bibitem[\protect\citeauthoryear{{Ade} et~al.,}{{Ade} et~al.}{2019}]{SO2019}
{Ade} P.,  et~al., 2019, \mn@doi [\jcap] {10.1088/1475-7516/2019/02/056}, \href
  {https://ui.adsabs.harvard.edu/abs/2019JCAP...02..056A} {2019, 056}

\bibitem[\protect\citeauthoryear{{Alam} et~al.,}{{Alam}
  et~al.}{2016}]{Boss2016combined}
{Alam} S.,  et~al., 2016, preprint, \href
  {http://adsabs.harvard.edu/abs/2016arXiv160703155A} {} (\mn@eprint {arXiv}
  {1607.03155})

\bibitem[\protect\citeauthoryear{{Alonso}, {Sanchez}, {Slosar}  \& {LSST Dark
  Energy Science Collaboration}}{{Alonso} et~al.}{2019}]{Alonso2019}
{Alonso} D.,  {Sanchez} J.,  {Slosar} A.,   {LSST Dark Energy Science
  Collaboration} 2019, \mn@doi [\mnras] {10.1093/mnras/stz093}, \href
  {https://ui.adsabs.harvard.edu/abs/2019MNRAS.484.4127A} {484, 4127}

\bibitem[\protect\citeauthoryear{{Baddour}}{{Baddour}}{2014}]{Baddour2014}
{Baddour} N.,  2014, \mn@doi [SpringerPlus] {10.1186/2193-1801-3-246}, \href
  {https://pubmed.ncbi.nlm.nih.gov/24877034/} {3, 246}

\bibitem[\protect\citeauthoryear{{Baldauf}, {Smith}, {Seljak}  \&
  {Mandelbaum}}{{Baldauf} et~al.}{2010}]{Baldauf2010}
{Baldauf} T.,  {Smith} R.~E.,  {Seljak} U.,   {Mandelbaum} R.,  2010, \mn@doi
  [\prd] {10.1103/PhysRevD.81.063531}, \href
  {http://adsabs.harvard.edu/abs/2010PhRvD..81f3531B} {81, 063531}

\bibitem[\protect\citeauthoryear{{Bohm} \& {Zech}}{{Bohm} \&
  {Zech}}{2014}]{Bohm2014}
{Bohm} G.,  {Zech} G.,  2014, \mn@doi [Nuclear Instruments and Methods in
  Physics Research A] {10.1016/j.nima.2014.02.021}, \href
  {https://ui.adsabs.harvard.edu/abs/2014NIMPA.748....1B} {748, 1}

\bibitem[\protect\citeauthoryear{{Bridle}, {Crittenden}, {Melchiorri},
  {Hobson}, {Kneissl}  \& {Lasenby}}{{Bridle} et~al.}{2002}]{Bridle2002}
{Bridle} S.~L.,  {Crittenden} R.,  {Melchiorri} A.,  {Hobson} M.~P.,  {Kneissl}
  R.,   {Lasenby} A.~N.,  2002, \mn@doi [\mnras]
  {10.1046/j.1365-8711.2002.05709.x}, \href
  {https://ui.adsabs.harvard.edu/abs/2002MNRAS.335.1193B} {335, 1193}

\bibitem[\protect\citeauthoryear{{DES Collaboration} et~al.,}{{DES
  Collaboration} et~al.}{2017}]{DES2017comb}
{DES Collaboration} et~al., 2017, preprint, \href
  {http://adsabs.harvard.edu/abs/2017arXiv170801530D} {} (\mn@eprint {arXiv}
  {1708.01530})

\bibitem[\protect\citeauthoryear{{Di Valentino} et~al.,}{{Di Valentino}
  et~al.}{2020}]{Valentino2020}
{Di Valentino} E.,  et~al., 2020, arXiv e-prints, \href
  {https://ui.adsabs.harvard.edu/abs/2020arXiv200811284D} {p. arXiv:2008.11284}

\bibitem[\protect\citeauthoryear{{Dor{\'e}} et~al.,}{{Dor{\'e}}
  et~al.}{2014}]{Spherex2014}
{Dor{\'e}} O.,  et~al., 2014, arXiv e-prints, \href
  {https://ui.adsabs.harvard.edu/abs/2014arXiv1412.4872D} {p. arXiv:1412.4872}

\bibitem[\protect\citeauthoryear{{Dore} et~al.,}{{Dore}
  et~al.}{2019}]{Wfirst2019}
{Dore} O.,  et~al., 2019, \baas, \href
  {https://ui.adsabs.harvard.edu/abs/2019BAAS...51c.341D} {51, 341}

\bibitem[\protect\citeauthoryear{{Doux} et~al.,}{{Doux}
  et~al.}{2021}]{Doux2021}
{Doux} C.,  et~al., 2021, \mn@doi [\mnras] {10.1093/mnras/stab661}, \href
  {https://ui.adsabs.harvard.edu/abs/2021MNRAS.tmp..712D} {}

\bibitem[\protect\citeauthoryear{{Efstathiou}}{{Efstathiou}}{2004}]{Efstathiou2004}
{Efstathiou} G.,  2004, \mn@doi [\mnras] {10.1111/j.1365-2966.2004.07530.x},
  \href {https://ui.adsabs.harvard.edu/abs/2004MNRAS.349..603E} {349, 603}

\bibitem[\protect\citeauthoryear{{Elsner}, {Leistedt}  \& {Peiris}}{{Elsner}
  et~al.}{2017}]{Elsner2017}
{Elsner} F.,  {Leistedt} B.,   {Peiris} H.~V.,  2017, \mn@doi [\mnras]
  {10.1093/mnras/stw2752}, \href
  {https://ui.adsabs.harvard.edu/abs/2017MNRAS.465.1847E} {465, 1847}

\bibitem[\protect\citeauthoryear{{Elvin-Poole} et~al.,}{{Elvin-Poole}
  et~al.}{2018}]{Elvin-Poole2018}
{Elvin-Poole} J.,  et~al., 2018, \mn@doi [\prd] {10.1103/PhysRevD.98.042006},
  \href {https://ui.adsabs.harvard.edu/abs/2018PhRvD..98d2006E} {98, 042006}

\bibitem[\protect\citeauthoryear{{Everett} et~al.,}{{Everett}
  et~al.}{2020}]{Everett2020}
{Everett} S.,  et~al., 2020, arXiv e-prints, \href
  {https://ui.adsabs.harvard.edu/abs/2020arXiv201212825E} {p. arXiv:2012.12825}

\bibitem[\protect\citeauthoryear{{Feldman}, {Kaiser}  \& {Peacock}}{{Feldman}
  et~al.}{1994}]{FKP}
{Feldman} H.~A.,  {Kaiser} N.,   {Peacock} J.~A.,  1994, \mn@doi [\apj]
  {10.1086/174036}, \href
  {https://ui.adsabs.harvard.edu/abs/1994ApJ...426...23F} {426, 23}

\bibitem[\protect\citeauthoryear{{G{\'o}rski}, {Hivon}, {Banday}, {Wandelt},
  {Hansen}, {Reinecke}  \& {Bartelmann}}{{G{\'o}rski}
  et~al.}{2005}]{Gorski2005}
{G{\'o}rski} K.~M.,  {Hivon} E.,  {Banday} A.~J.,  {Wandelt} B.~D.,  {Hansen}
  F.~K.,  {Reinecke} M.,   {Bartelmann} M.,  2005, \mn@doi [\apj]
  {10.1086/427976}, \href {http://adsabs.harvard.edu/abs/2005ApJ...622..759G}
  {622, 759}

\bibitem[\protect\citeauthoryear{{Hamana} et~al.,}{{Hamana}
  et~al.}{2020}]{Hamana2020}
{Hamana} T.,  et~al., 2020, \mn@doi [\pasj] {10.1093/pasj/psz138}, \href
  {https://ui.adsabs.harvard.edu/abs/2020PASJ...72...16H} {72, 16}

\bibitem[\protect\citeauthoryear{{Hamilton}}{{Hamilton}}{1997}]{Hamilton1997}
{Hamilton} A.~J.~S.,  1997, \mn@doi [\mnras] {10.1093/mnras/289.2.285}, \href
  {https://ui.adsabs.harvard.edu/abs/1997MNRAS.289..285H} {289, 285}

\bibitem[\protect\citeauthoryear{{Heymans} et~al.,}{{Heymans}
  et~al.}{2021}]{Heymans2021}
{Heymans} C.,  et~al., 2021, \mn@doi [\aap] {10.1051/0004-6361/202039063},
  \href {https://ui.adsabs.harvard.edu/abs/2021A&A...646A.140H} {646, A140}

\bibitem[\protect\citeauthoryear{{Hivon}, {G{\'o}rski}, {Netterfield}, {Crill},
  {Prunet}  \& {Hansen}}{{Hivon} et~al.}{2002}]{Hivon2002}
{Hivon} E.,  {G{\'o}rski} K.~M.,  {Netterfield} C.~B.,  {Crill} B.~P.,
  {Prunet} S.,   {Hansen} F.,  2002, \mn@doi [\apj] {10.1086/338126}, \href
  {https://ui.adsabs.harvard.edu/abs/2002ApJ...567....2H} {567, 2}

\bibitem[\protect\citeauthoryear{{Joachimi} et~al.,}{{Joachimi}
  et~al.}{2021}]{Joachimi2021}
{Joachimi} B.,  et~al., 2021, \mn@doi [\aap] {10.1051/0004-6361/202038831},
  \href {https://ui.adsabs.harvard.edu/abs/2021A&A...646A.129J} {646, A129}

\bibitem[\protect\citeauthoryear{{LSST Dark Energy Science
  Collaboration}}{{LSST Dark Energy Science Collaboration}}{2012}]{LSST_DESC}
{LSST Dark Energy Science Collaboration} 2012, arXiv e-prints, \href
  {https://ui.adsabs.harvard.edu/abs/2012arXiv1211.0310L} {p. arXiv:1211.0310}

\bibitem[\protect\citeauthoryear{{Landy} \& {Szalay}}{{Landy} \&
  {Szalay}}{1993}]{Landy1993}
{Landy} S.~D.,  {Szalay} A.~S.,  1993, \mn@doi [\apj] {10.1086/172900}, \href
  {http://adsabs.harvard.edu/abs/1993ApJ...412...64L} {412, 64}

\bibitem[\protect\citeauthoryear{{Lange}, {Leauthaud}, {Singh}, {Guo}, {Zhou},
  {Smith}  \& {Cyr-Racine}}{{Lange} et~al.}{2021}]{Lange2021}
{Lange} J.~U.,  {Leauthaud} A.,  {Singh} S.,  {Guo} H.,  {Zhou} R.,  {Smith}
  T.~L.,   {Cyr-Racine} F.-Y.,  2021, \mn@doi [\mnras] {10.1093/mnras/stab189},
  \href {https://ui.adsabs.harvard.edu/abs/2021MNRAS.502.2074L} {502, 2074}

\bibitem[\protect\citeauthoryear{{Leistedt}, {Peiris}, {Mortlock},
  {Benoit-L{\'e}vy}  \& {Pontzen}}{{Leistedt} et~al.}{2013}]{Leistedt2013}
{Leistedt} B.,  {Peiris} H.~V.,  {Mortlock} D.~J.,  {Benoit-L{\'e}vy} A.,
  {Pontzen} A.,  2013, \mn@doi [\mnras] {10.1093/mnras/stt1359}, \href
  {https://ui.adsabs.harvard.edu/abs/2013MNRAS.435.1857L} {435, 1857}

\bibitem[\protect\citeauthoryear{{Levi} et~al.,}{{Levi}
  et~al.}{2019}]{DESI2019}
{Levi} M.,  et~al., 2019, in Bulletin of the American Astronomical Society.
  p.~57 (\mn@eprint {arXiv} {1907.10688})

\bibitem[\protect\citeauthoryear{{Li}, {Singh}, {Yu}, {Feng}  \& {Seljak}}{{Li}
  et~al.}{2019}]{Li2019}
{Li} Y.,  {Singh} S.,  {Yu} B.,  {Feng} Y.,   {Seljak} U.,  2019, \mn@doi
  [\jcap] {10.1088/1475-7516/2019/01/016}, \href
  {https://ui.adsabs.harvard.edu/abs/2019JCAP...01..016L} {2019, 016}

\bibitem[\protect\citeauthoryear{{Neveux} et~al.,}{{Neveux}
  et~al.}{2020}]{Neveux2020}
{Neveux} R.,  et~al., 2020, \mn@doi [\mnras] {10.1093/mnras/staa2780}, \href
  {https://ui.adsabs.harvard.edu/abs/2020MNRAS.499..210N} {499, 210}

\bibitem[\protect\citeauthoryear{{Ng} \& {Liu}}{{Ng} \& {Liu}}{1999}]{Ng1999}
{Ng} K.-W.,  {Liu} G.-C.,  1999, \mn@doi [International Journal of Modern
  Physics D] {10.1142/S0218271899000079}, \href
  {https://ui.adsabs.harvard.edu/abs/1999IJMPD...8...61N} {8, 61}

\bibitem[\protect\citeauthoryear{{Planck Collaboration} et~al.,}{{Planck
  Collaboration} et~al.}{2020}]{Planck2018lensing}
{Planck Collaboration} et~al., 2020, \mn@doi [\aap]
  {10.1051/0004-6361/201833886}, \href
  {https://ui.adsabs.harvard.edu/abs/2020A&A...641A...8P} {641, A8}

\bibitem[\protect\citeauthoryear{{Rezaie}, {Seo}, {Ross}  \&
  {Bunescu}}{{Rezaie} et~al.}{2020}]{Rezaie2020}
{Rezaie} M.,  {Seo} H.-J.,  {Ross} A.~J.,   {Bunescu} R.~C.,  2020, \mn@doi
  [\mnras] {10.1093/mnras/staa1231}, \href
  {https://ui.adsabs.harvard.edu/abs/2020MNRAS.495.1613R} {495, 1613}

\bibitem[\protect\citeauthoryear{{Ross} et~al.,}{{Ross}
  et~al.}{2012}]{Ross2012}
{Ross} A.~J.,  et~al., 2012, \mn@doi [\mnras]
  {10.1111/j.1365-2966.2012.21235.x}, \href
  {http://adsabs.harvard.edu/abs/2012MNRAS.424..564R} {424, 564}

\bibitem[\protect\citeauthoryear{{Ross} et~al.,}{{Ross}
  et~al.}{2020}]{Ross2020}
{Ross} A.~J.,  et~al., 2020, \mn@doi [\mnras] {10.1093/mnras/staa2416}, \href
  {https://ui.adsabs.harvard.edu/abs/2020MNRAS.498.2354R} {498, 2354}

\bibitem[\protect\citeauthoryear{{Schaan} \& {White}}{{Schaan} \&
  {White}}{2021}]{Schaan2021}
{Schaan} E.,  {White} M.,  2021, arXiv e-prints, \href
  {https://ui.adsabs.harvard.edu/abs/2021arXiv210301971S} {p. arXiv:2103.01971}

\bibitem[\protect\citeauthoryear{{Schneider}, {Eifler}  \&
  {Krause}}{{Schneider} et~al.}{2010}]{Schneider2010}
{Schneider} P.,  {Eifler} T.,   {Krause} E.,  2010, \mn@doi [\aap]
  {10.1051/0004-6361/201014235}, \href
  {https://ui.adsabs.harvard.edu/abs/2010A&A...520A.116S} {520, A116}

\bibitem[\protect\citeauthoryear{{Singh}, {Mandelbaum}, {Seljak}, {Slosar}  \&
  {Vazquez Gonzalez}}{{Singh} et~al.}{2017}]{Singh2017cov}
{Singh} S.,  {Mandelbaum} R.,  {Seljak} U.,  {Slosar} A.,   {Vazquez Gonzalez}
  J.,  2017, \mn@doi [\mnras] {10.1093/mnras/stx1828}, \href
  {http://adsabs.harvard.edu/abs/2017MNRAS.471.3827S} {471, 3827}

\bibitem[\protect\citeauthoryear{{Singh}, {Mandelbaum}, {Seljak},
  {Rodr{\'\i}guez-Torres}  \& {Slosar}}{{Singh} et~al.}{2020}]{Singh2020cosmo}
{Singh} S.,  {Mandelbaum} R.,  {Seljak} U.,  {Rodr{\'\i}guez-Torres} S.,
  {Slosar} A.,  2020, \mn@doi [\mnras] {10.1093/mnras/stz2922}, \href
  {https://ui.adsabs.harvard.edu/abs/2020MNRAS.491...51S} {491, 51}

\bibitem[\protect\citeauthoryear{Slosar, Seljak  \& Makarov}{Slosar
  et~al.}{2004}]{Slosar2004}
Slosar A.,  Seljak U.,   Makarov A.,  2004, \mn@doi [Phys. Rev. D]
  {10.1103/PhysRevD.69.123003}, 69, 123003

\bibitem[\protect\citeauthoryear{{Wandelt}, {Hivon}  \& {G{\'o}rski}}{{Wandelt}
  et~al.}{2001}]{Wandelt2001}
{Wandelt} B.~D.,  {Hivon} E.,   {G{\'o}rski} K.~M.,  2001, \mn@doi [\prd]
  {10.1103/PhysRevD.64.083003}, \href
  {https://ui.adsabs.harvard.edu/abs/2001PhRvD..64h3003W} {64, 083003}

\bibitem[\protect\citeauthoryear{{Weinberg}, {Mortonson}, {Eisenstein},
  {Hirata}, {Riess}  \& {Rozo}}{{Weinberg} et~al.}{2013}]{Weinberg2013}
{Weinberg} D.~H.,  {Mortonson} M.~J.,  {Eisenstein} D.~J.,  {Hirata} C.,
  {Riess} A.~G.,   {Rozo} E.,  2013, \mn@doi [\physrep]
  {10.1016/j.physrep.2013.05.001}, \href
  {http://adsabs.harvard.edu/abs/2013PhR...530...87W} {530, 87}

\makeatother
\end{thebibliography}

\onecolumn
\appendix
%
%

	\section{Comparison of window weighting on noise}
	\label{app:noise_inverse_dist}
		In this appendix we prove the claim from section~\ref{ssec:shot_noise} that $\widebar{\left[\frac{1}{W(\vx)}\right]}\geq\widebar{W}(\vx)$.
		
		We begin by noting that by $W(\vx)\in[0,\infty)$ and $\widebar{W}(\vx)=1$. For the case $W(\vx)\in[0,\infty)$, proof is trivial as $\widebar{\left[\frac{1}{W(\vx)}\right]}\rightarrow\infty$. Hence we will focus on the
		case $W(\vx)\in(0,\infty)$ (i.e. 0 is excluded).
		
		We can write $W(x)$ in terms of a mean zero variable as 
		\begin{align}
			W(\vx)&=1+w(\vx),
		\end{align}
		where 
		\begin{align}
			\widebar{W}(\vx)&=1+\widebar{w}(\vx)=1, \hspace{20pt}\widebar{w}(\vx)=0.
		\end{align}
		The mean of $1/W(\vx)$ is then
		\begin{align}
			\widebar {\left[\frac{1}{{W_g(\vx)}}\right]}&=\widebar{\left[{\frac{1}{1+w(\vx)}}\right]}.
		\end{align}
		Defining $y= w(\vx)$ and rewriting we get
		\begin{align}
			\widebar {\left[\frac{1}{{W_g(\vx)}}\right]}&=\int_{-1}^{\infty}\frac{1}{1+y}P(y)dy=\int_{-1}^{1}\frac{1}{1+y}P(y)dy+\int_{1}^{\infty}\frac{1}{1+y}P(y)dy.
		\end{align}
		$P(y)$ is the probability distribution and 
		in the second step we split the integral into two ranges, $y\in(-1,1)$ and $y\in[1,\infty)$. Using the Taylor series in the first integral, we get
		\begin{align}
			\widebar {\left[\frac{1}{{W_g(\vx)}}\right]}&=\int_{-1}^{1}\left[1+\sum_i (-1)^i y^i\right]P(y)dy+\int_{1}^{\infty}\frac{1}{1+y}P(y)dy\\
									&=\int_{-1}^{1}dyP(y)(1-y)+ \int_{-1}^{1}dyP(y) ([y^2-y^3]+[y^4-y^5]\dots)+\int_{1}^{\infty}\frac{1}{1+y}P(y)dy.\\
		\end{align}
		Now we use the fact that $\mean{y}=0$ and $\int_{-1}^{\infty}P(y)=1$, to change the limits of the first integral from $y\in(-1,1)$ to $y\in[1,\infty)$
		\begin{align}
			\widebar {\left[\frac{1}{{W_g(\vx)}}\right]}&=1+\int_{1}^{\infty}dyP(y)(-1+y)+ \int_{-1}^{1}dyP(y) ([y^2-y^3]+[y^4-y^5]\dots)+\int_{1}^{\infty}\frac{1}{1+y}P(y)dy\\
									&=1+\int_{-1}^{1}dyP(y) ([y^2-y^3]+[y^4-y^5]\dots)+\int_{1}^{\infty}\frac{y^2}{1+y}P(y)dy\geq1.
		\end{align}
		Notice that the quantities inside square brackets, $[]$, are always positive and
		the last integral is also non negative, thus proving that $\widebar {\left[\frac{1}{{W_g(\vx)}}\right]}\geq1$

\section{More general weighting}
	\label{appn:general_weighting}
	\refereetwo{In this appendix we derive the window and the noise effects in estimator of eq.~\eqref{eq:Pdelta_g} when more general weighting schemes are used.}
	
	Typically in a LSS survey, the galaxies are assigned weights which may depend on some intrinsic property of galaxy, we call such weights $w_{0,i}$ and another weight dependent on the variance, $w_{v,i}$, 
	such that the total weight is given by
	\begin{align}
		w_i=w_{0,i}w_{v,i}
	\end{align}
	The inverse variance weight is usually written as 
	\begin{align}
		w_{v,i}=\frac{N}{\text{Var}_i}=\frac{N}{C+\sigma^2_{m,i}},
	\end{align}
	where $\sigma^2_{m,i}$ is contributed by the measurement noise and $C$ is the sampling noise in galaxy field. $C=1$ for galaxies and $C=\sigma_e^2$, i.e. shape noise, for shear. $N$ is the normalization
	of the weights.
	The observed effective number of galaxies in a pixel are then
	\begin{align}
		{n}_g(\vx)=\sum_i w_i =\mean{n_g(\vx)}\widebar{w}_i(\vx)(1+\delta(\vx)).
	\end{align}
	$\widebar{w}_i(\vx)$ is mean of weights within the pixel. The windowed overdensity field is 
	\begin{align}
		{\delta}_{g,W}(\vx)=\frac{{n_g(\vx)}}{\widebar{n}_g\widebar w_i}-\frac{\mean{n_g(\vx)}}{\widebar{n}_g\widebar w_i}.
	\end{align}
	$\widebar w_i$ is the sample mean of all the weights and is usually normalized to be 1. The window in this case is,
	\begin{align}
		W_g(\vx)=\frac{\mean{n_g(\vx)}\widebar{w}_i(\vx)}{\widebar{n}_g\widebar w_i}.
	\end{align}
	The variance is given as 
	\begin{align}
		\delta_N^2(\vx)=\frac{1}{\widebar n_g^2\widebar w_i^2}\sum_i w_i^2 \text{Var}_i=\frac{N}{\widebar n_g^2\widebar w_i^2}\sum_i w_{0,i}^2 w_{v,i}.
	\end{align}
	Assuming that $w_{0,i}$ and $w_{v,i}$ are uncorrelated, we can write the sum as 	
	\begin{align}
		\sum_i w_{0,i}^2 w_{v,i}=\mean{n_g(\vx)} w_{v,i}(\vx)w^2_{0,i}(\vx),
	\end{align}
	to obtain
	\begin{align}
		\mean{\delta_N^2(\vx)}=N \frac{{\mean{n_g(\vx)} w_{v,i}(\vx)w^2_{0,i}}(\vx)}{\widebar n_g^2\widebar w_i^2}=
			N \frac{W_g(\vx)}{\widebar n_g\widebar w_i}\frac{w^2_{0,i}(\vx)}{w_{0,i}(\vx)}.
			\label{eq:gen_noise_window}
	\end{align}
	Averaging over the survey we get
	\begin{align}
		\mean{\delta_N^2}=N
			\frac{\widebar{W}_g}{\widebar n_g \widebar{w}_{v,i}}\frac{\widebar{w^2_{0,i}}}{\widebar w_{0,i}\widebar w_{0,i}}.
			\label{eq:gen_noise_window}
	\end{align}
	 Sometimes shot noise is described in terms of effective number density of galaxies, where
	\begin{align}
		\frac{1}{n_g^{eff}}=\frac{1}{\widebar n_g}\frac{\widebar{w^2}_{0,i}}{\widebar{w}_{0,i}\widebar w_{0,i}}.
	\end{align}
		Notice that in the absence of $w_{0,i}$ eq.~\eqref{eq:gen_noise_window} is equivalent to eq.~\eqref{eq:N_g2} (in eq.~\eqref{eq:N_g2} $N=1$). 
		
		\refereetwo{In general the additional weights do change the window as well as the dependence of noise in the window. Since we only worked with shot noise in this paper and 
		subtracted out the correct noise from pseudo-$C_\ell$ measurements, this does not affect the results in the main part of this paper. However, these results highlight the dependence of 
		noise on the window and weighting and such dependencies will need to be carefully modeled both for noise modeling and covariance matrix calculations. }
	
\section{Window coupling matrix}
	\label{app:Ang_PS}
	\refereetwo{In this appendix we derive the response of the pseudo-$C_\ell$ power spectra to window power spectra.}
	
	From \cite{Hivon2002}, the pseudo-$C_\ell$ power spectra is given as
	\begin{align}
        & D_\ell=
            \sum_{\ell'}C_{\ell'}{\frac{(2\ell'+1)}{4\pi}}\sum_{\ell''}
             W_{\ell''}(2\ell''+1)
            \wj{\ell}{\ell'}{\ell''}{s_1}{-s_1}{0}\wj{\ell}{\ell'}{\ell''}{s_2}{-s_2}{0}\label{aeq:Dl},\\
            D_\ell&=M_{\ell,\ell'}C_{\ell'},
    \end{align}
	where in second equation we used the definition of the coupling matrix from eq.~\eqref{eq:coupling_M}.
	
	In eq.~\eqref{aeq:Dl}, we can switch the order of summation over $\ell'$ and $\ell''$ to write
    \begin{align}
        & D_\ell=
        \sum_{\ell''}W_{\ell''}\frac{(2\ell''+1)}{4\pi}
            \sum_{\ell'}C_{\ell'}{(2\ell'+1)}
            \wj{\ell}{\ell'}{\ell''}{s_1}{-s_1}{0}\wj{\ell}{\ell'}{\ell''}{s_2}{-s_2}{0}\label{aeq:Dl},\\
        &D_\ell=M^W_{\ell,\ell''}W_{\ell''},\label{aeq:Dl_W}
	\end{align}
	where 
    \begin{align}
            M^W_{\ell,\ell''}={\frac{(2\ell''+1)}{4\pi}}\sum_{\ell'}
             C_{\ell'}(2\ell'+1)
            \wj{\ell}{\ell'}{\ell''}{s_1}{-s_1}{0}\wj{\ell}{\ell'}{\ell''}{s_2}{-s_2}{0}.
    \end{align}
	We can also use the symmetries of wigner-$3j$ symbols to change the ordering of $\ell',\ell''$ if desired.
	
	\refereetwo{From eq.~\eqref{aeq:Dl} and \eqref{aeq:Dl_W}, we notice that the the pseudo-$C_\ell$ power spectra is symmetric in its response to the window and the underlying density field we 
	wish to study. This highlights the importance of modeling window properly and we used the eq.~\eqref{aeq:Dl_W} to study the impact the multiplicative biases in window (biased $W_{\ell''}$)
	have on $D_\ell$.}

	\section{Correlation functions}
		\label{app:xi}
		Here we derive the expressions for correlation functions in the presence of a survey window. We closely follow the expressions in \cite{Ng1999}.
		
		We begin by writing the two point correlation function in presence of windows as 
		\begin{align}
          	\mean{\delta_1\delta_2}(\vtheta)=&\frac{1}{\xi_W(\vtheta)}\int d^2\vtheta'\delta_1(\vtheta')
                  \delta_2(\vtheta'+\vtheta)W_1(\vtheta')W_2(\vtheta'+\vtheta)\\
                               =&\frac{1}{\xi_W(\vtheta)}\int d^2\vtheta'\sum_{l_{1-4},m_{1-4}}\delta_{1,l_1,m_1}\delta_{2,l_2,m_2}
                  W_1(l_3,m_3)W_2(l_4,m_4)_{s_1}Y_{\ell_1,m_1}(\vtheta')_{s_2}Y_{l_2,m_2}(\vtheta'+\vtheta)Y_{l_3,m_3}
                  (\vtheta')Y_{l_4,m_4}(\vtheta'+\vtheta),
        \end{align}
        where in the second equation we simply wrote the $\delta_i$ and $W_i$ in terms of their spherical harmonic transforms (equivalent of Fourier Transform on a sphere).
        
        Noting that $\delta_{1,l_1,m_1}\delta_{2,l_2,m_2}=C_{\ell_1}\delta_D(l_1,l_2)\delta_D(m_1,m_2)$, we get
        \begin{align}
             \mean{\delta_1\delta_2}(\vtheta)=&\frac{1}{\xi_W(\vtheta)}\int d^2\vtheta'\sum_{l_{1-4},m_{1-4}}C_{\ell_1}\delta_D(l_1,l_2)\delta_D(m_1,m_2)
                  W_1(l_3,m_3)W_2(l_4,m_4) _{s_1}Y_{\ell_1,m_1}(\vtheta')_{s_2}Y_{l_2,m_2}(\vtheta'+\vtheta)Y_{l_3,m_3}
                  (\vtheta')Y_{l_4,m_4}(\vtheta'+\vtheta),\\
             \mean{\delta_1\delta_2}(\vtheta)=&\frac{1}{\xi_W(\vtheta)}\int d^2\vtheta'\sum_{l_{1,3,4},m_{1,3,4}}C_{\ell_1}
                  W_1(l_3,m_3)W_2(l_4,m_4) _{s_1}Y_{\ell_1,m_1}(\vtheta')_{s_2}Y_{\ell_1,m_1}(\vtheta'+\vtheta)Y_{l_3,m_3}
                  (\vtheta')Y_{l_4,m_4}(\vtheta'+\vtheta).
        \end{align}
        Where in the second step I carried out the sums over $\delta_D$. Now we use the spherical harmonics identity \citep{Ng1999}, 
        $\sum_m {_{s_1}Y_{\ell_1,m_1}}(\vtheta')_{s_2}Y_{\ell_1,m_1}(\vtheta'+\vtheta)=\sqrt{\frac{2\ell_1+1}{4\pi}}(-1)^{s1-s2}{_{s_1}Y_{\ell_1,s_2}}(\vtheta)$ to obtain
        \begin{align}
             \mean{\delta_1\delta_2}(\vtheta)=&\frac{1}{\xi_W(\vtheta)}\int d^2\vtheta'\sum_{l_{1,3,4},m_{3,4}}C_{\ell_1}
                  W_1(l_3,m_3)W_2(l_4,m_4){_{-s_1}}Y_{\ell_1,s_2}(\vtheta)Y_{l_3,m_3}(\vtheta')Y_{l_4,m_4}(\vtheta'+\vtheta),
        \end{align}
        where I omitted the $(-1)^{s1-s2}$ factor since $s_1$ and $s_2$ are even numbers for the tracers we use. Rearranging the terms, we get
        \begin{align}
             \mean{\delta_1\delta_2}(\vtheta)=&\frac{1}{\xi_W(\vtheta)}\sum_{\ell_1}\sqrt{\frac{2\ell_1+1}{4\pi}} {_{-s_1}}Y_{\ell_1,s_2}(\vtheta)C_{\ell_1} \int d^2\vtheta'\sum_{l_{3,4},m_{3,4}}
                  W_1(l_3,m_3)W_2(l_4,m_4)Y_{l_3,m_3}(\vtheta')Y_{l_4,m_4}(\vtheta'+\vtheta)\label{eq:xi_W_lm}.\\
    		=&\frac{1}{\xi_W(\vtheta)}\sum_{\ell_1} \sqrt{\frac{2\ell_1+1}{4\pi}} {_{-s_1}}Y_{\ell_1,s_2}(\vtheta)C_{\ell_1} \xi_W(\vtheta).
      \end{align}
      The term inside the integral is simply the correlation function of the window, denoted by $\xi_W(\vtheta)$. 
       
      Using 
      \begin{align}
      {_{-s_1}}Y_{\ell_1,s_2}(\theta,\phi)=\sqrt{\frac{2\ell_1+1}{4\pi}}{_{-s_1}}d_{\ell_1,s_2}(\cos(\theta))e^{-is\phi}, 
      \end{align}
       and carrying out the angular integrals \citep[see discussion in ][]{Ng1999}, we obtain
    	\begin{align}
          	\mean{\delta_1\delta_2}(\theta)=&\frac{\xi_W(\theta)}{\xi_W(\theta)}\sum_{\ell_1} \frac{2\ell+1}{4\pi}{_{s_1}}d_{\ell_1,s_2}(\cos\theta)C_{\ell_1} .
			\label{eq:xi_win_app}
		\end{align}  
		We used the full sky averaging expressions from eq. 7.3 of  \cite{Ng1999}, to simplify the window correlation function, i.e. 
      \begin{align}
      	\xi_W(\vtheta)=\sum_{\ell_1} \sqrt{\frac{2\ell_1+1}{4\pi}} {_{0}}Y_{\ell,0}(\theta)W_{\ell}=\sum_{\ell_1} {\frac{2\ell_1+1}{4\pi}} {_{0}}d_{\ell,0}(\cos(\theta))W_{\ell} = \xi_W(\theta).
      \end{align}
      $W_{\ell}$ is the pseudo-$C_\ell$ power spectra of the window. 

      \refereetwo{While the window effect appears to cancel in eq.~\eqref{eq:xi_win_app}, as discussed in section~\ref{ssec:LS_estimator}, the binning operator act separately on numerator and denominator, 
      hence the window effects do not fully cancel. Therefore the correlation functions are still sensitive to the window and here we have shown that the window effects separate out from the 
      density field such that the estimator is only sensitive to the correlation function of the window.}
      
      \section{Scale cuts}
      	\label{app:corr_scale_cuts}
      	\subsection{$D_\ell$ reconstruction from $\xi$}	
			\label{app:cl_xi_scale_cuts}
		\refereetwo{Here we show that the inverse Hankel transform of correlation functions measured over limited range of scales leads to pseudo-$C_\ell$ power spectra. This proof also shows the 
		equivalence of correlation function and pseudo-$C_\ell$ power spectra when window correlation function $\xi_W(\theta)$ is used in place of $w(\theta)$ in the equations below.}
		
      	For scale cuts on correlation functions, we have
		\begin{align}
			D_\ell&=2\pi\int d\theta\sin(\theta)\xi(\theta)w(\theta)\sqrt{\frac{4\pi}{2\ell+1}}{_{-s_1}}Y_{\ell,s_2}(\theta).
		\end{align}
		Writing $\xi(\theta)$ and $w(\theta)$ in terms of their Fourier counterparts, we get
		\begin{align}
				 D_\ell&=\sum_{\ell_1,\ell_2} C_{\ell_1}w_{\ell_2}\sqrt{\frac{2\ell_1+1}{4\pi}}\sqrt{\frac{2\ell_2+1}{4\pi}}	\int d\theta\sin(\theta){_{-s_1}}Y_{\ell,s_2}(\theta){_{-s_1}}
				 Y_{\ell_1,s_2}(\theta){_{-s_{1w}}}Y_{\ell_2,s_{2w}}(\theta),\\
				 D_\ell&=\sum_{\ell_1,\ell_2} C_{\ell_1}w_{\ell_2}\sqrt{\frac{4\pi}{2\ell+1}}{\frac{(2\ell_1+1)(2\ell_2+1)}{4\pi}}\sqrt{\frac{2\ell+1}{4\pi}}
				 \wj{\ell}{\ell_1}{\ell_2}{s_2}{-s_2}{s_{2w}}\wj{\ell}{\ell_1}{\ell_2}{s_1}{-s_1}{s_{1w}},\\
				 D_\ell&=\sum_{\ell_1,\ell_2} C_{\ell_1}w_{\ell_2}{\frac{(2\ell_1+1)(2\ell_2+1)}{4\pi}}
				 \wj{\ell}{\ell_1}{\ell_2}{s_2}{-s_2}{s_{2w}}\wj{\ell}{\ell_1}{\ell_2}{s_1}{-s_1}{s_{1w}},\label{aeq:xi_Dl}
		\end{align}  
		where in the second equation we wrote the integral over three spherical harmonics in terms of the wigner-3j function and the last equation is very similar to eq.~\eqref{aeq:Dl}, with
		$s_{1w}=s_{2w}=0$.
		As stated earlier, we can replace $w(\theta)$ with the correlation function of the window $\xi_W(\theta)$ in which case $w(\ell_2)$ gets replaced with the window power spectra $W(\ell_2)$ 
		and we obtain the expressions identical to eq.~\eqref{aeq:Dl}, thus showing that the power spectra and correlation function estimators are identical when full range of scales in $\ell,
		\theta$ is used.
		
		\subsection{Modeling $\xi$ with cuts on $C_\ell$}
			\refereetwo{Here we derive the expressions for convolution operator acting on the correlation functions to account for the model cuts in the Fourier space.}
			
			We wish to derive the method to correct $\xi$ for model cuts in the $\ell$ space, i.e.
			\begin{align}
				\xi_{cut}(\theta)&=\sum_{\ell} b_\ell C_{\ell} \sqrt{\frac{2\ell+1}{4\pi}} {_{s_1}Y_{\ell,{s_2}}}(\theta).
				\label{eqa:xi_c_l}
			\end{align}
			 $b_\ell$ is the function that applies the cuts in $\ell$.
			
			Following \cite{Baddour2014}, we make an ansatz that 
			\begin{align}
				\xi_{cut}(\theta)=\int_0^{2\pi} d\phi_1\int d\theta_1\sin(\theta_1)\xi(\theta_1)b(\theta,\theta_1),
				\label{eqa:xi_c0}
			\end{align}
			where
			\begin{align}
				b(\theta,\theta_1)=\sum_{\ell} b_\ell \frac{2\ell+1}{4\pi} {_{s_1}Y_{\ell,{s_2}}}(\theta)_{s_1}Y_{\ell,{s_2}}(\theta_1).
			\end{align}
			Plugging $b(\theta,\theta_1)$ back we get
			\begin{align}
				\xi_{cut}(\theta)&=\int d\phi_1\int d\theta_1\sin(\theta_1)\sum_{\ell} b_\ell \frac{2\ell+1}{4\pi} {_{s_1}Y_{\ell,{s_2}}}(\theta)_{s_1}Y_{\ell,{s_2}}(\theta_1)
								\sum_{\ell_1} C_{\ell_1} \sqrt{\frac{2\ell+1}{4\pi}} {_{s_1}Y_{\ell_1,{s_2}}}(\theta_1)_{s_1},\\
								&=\sum_{\ell} b_\ell C_{\ell} \sqrt{\frac{2\ell+1}{4\pi}} {_{s_1}Y_{\ell,{s_2}}}(\theta).
			\end{align}
			Hence showing that eq.~\eqref{eqa:xi_c0} is equivalent to eq.~\eqref{eqa:xi_c_l}.

	\section{Simulations}
	\label{app:sims}
		This appendix describes the simulations and the assumptions about data samples used in the examples presented in the paper.
		
		For shear, we used a LSST shear sample properties, with $\widebar{n}_g=26$ arcminutes$^{-2}$, $f_{sky}=0.3$, shape noise $\sigma_\gamma=0.26$ per component. All the galaxies are 
		assumed to be in a narrow redshift bin at $z=1$. For the galaxy sample, we used $\widebar{n}_g=10$ arcminutes$^{-2}$, $f_{sky}=0.3$, $b_g=1$. Magnification and intrinsic alignments
		are set to be zero.

			\begin{figure*}
			\begin{subfigure}[t]{.3\columnwidth}
    	    		\centering
        	 		\includegraphics[width=.85\columnwidth]{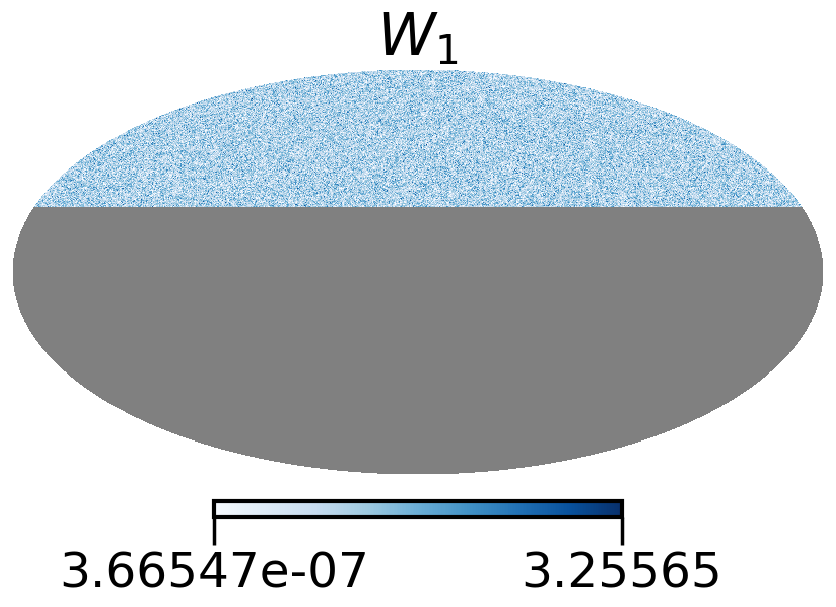}
					\caption{}
					\label{fig:W1}
			\end{subfigure}
			\begin{subfigure}[t]{.3\columnwidth}
    	    		\centering
        	 		\includegraphics[width=.85\columnwidth]{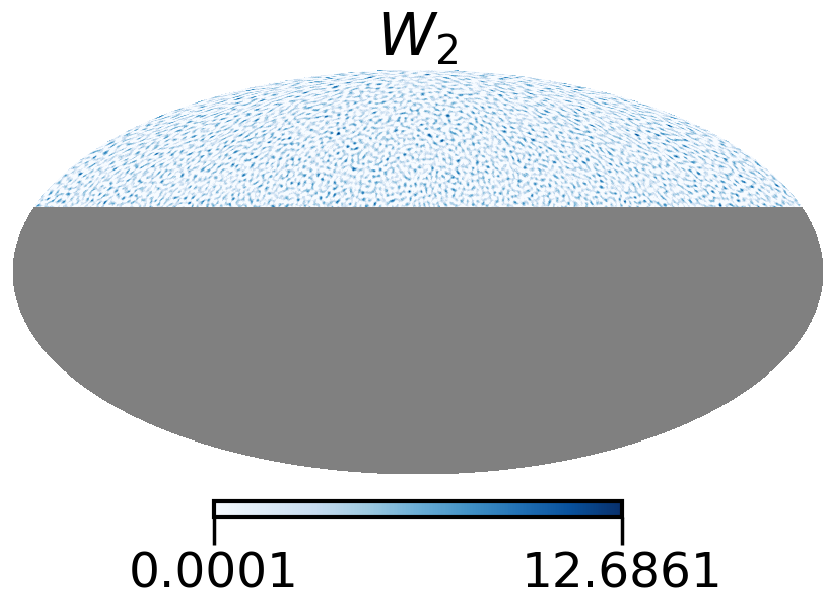}
					\caption{}
					\label{fig:W2}
			\end{subfigure}
			\begin{subfigure}[t]{.3\columnwidth}
    	    		\centering
        	 		\includegraphics[width=.85\columnwidth]{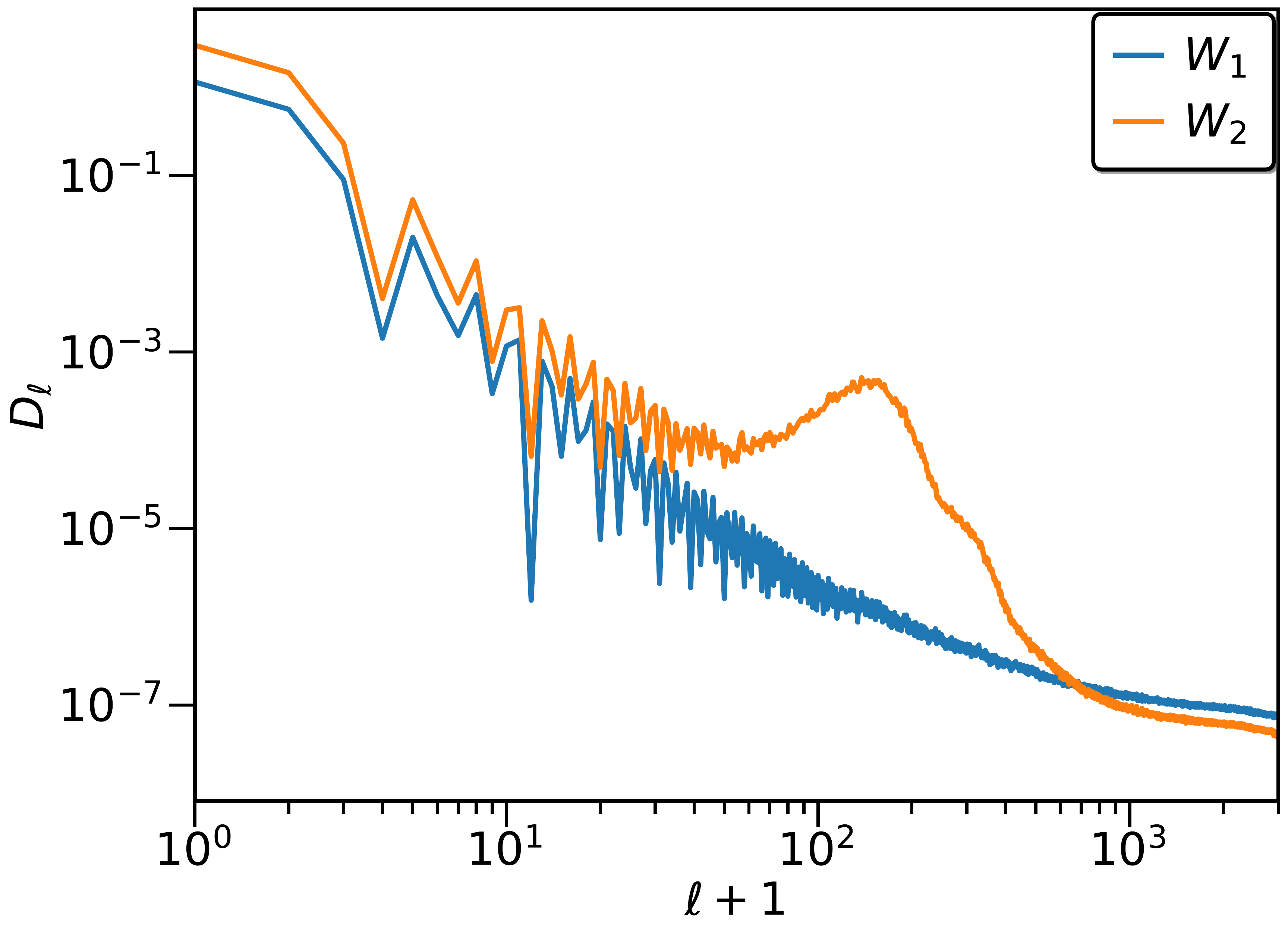}
					\caption{}
					\label{fig:W_cl}
			\end{subfigure}
			\caption{Window functions used for the simulations as well as other calculations using windows in the main part of the paper. $W_1$ is obtained using same $C_\ell$ as the galaxy 
			sample 
			($C_{\ell,gg}$) and represents a realistic window for the case galaxy shear. $W_2$ represents a more complex window with an additional gaussian power spectra 
			($C_{\ell,gg}+C_{\ell,\text{gaussian}}$), where $C_{\ell,\text{gaussian}}$ peaks at $\ell\sim 200$ and has the width of $\sigma\sim50$. Both windows have constants added such that the 
			minimum value is 0. c) The power spectra of two windows.
			}
			\label{fig:Window}
		\end{figure*}
	
		Fig.~\ref{fig:Window} shows the two different window functions used in the examples presented in main part of the paper.

\end{document}